\documentclass{article}
\usepackage{amsmath}
\usepackage{graphicx}
\usepackage{float}
\usepackage{caption}
\usepackage{subfigure}

\input{tcilatex}
\graphicspath{{"F:/Research/ResearchStudents/PostPhD/Azeem/Draft/Images/"}}

\begin{document}

\title{Study of rare mesonic decays involving di-neutrinos in their final
state }
\author{Azeem Mir\footnotemark , Farida\ Tahir\footnotemark[1], Shakeel
Mahmood\footnotemark , Shi- Hai Dong\footnotemark \newline
\\
%EndAName
(Physics Department, COMSATS Institute of Information Technology, Islamabad. 
\newline
\footnotemark[1]\\
Air University, PAF Complex, Service Road, E-9, Islamabad, Pakistan%
\footnotemark[2].\newline
\\
CIDETEC, Instituto Polit\'{e}cnico Nacional, Unidad Professional, Adolfo L%
\'{o}pez Mateos,\newline
\\
CDMX 07700, Mexico\footnotemark[3].)}
\date{}
\maketitle

\begin{abstract}
We have carried out phenomenological implication of R-parity violating ($\NEG%
{R}_{p}$) Minimal Supersymmetric Model (MSSM) via analyses of pure leptonic($%
M\rightarrow \nu \bar{\nu}$) and semileptonic decays of pseudo-scalar mesons(%
$M\rightarrow X\nu \bar{\nu}$). These analyses involve prediction of
branching fraction of pure leptonic decays by using experimental
limits/bounds derived from the study of semileptonic decays on $\NEG{R}_{p}$
parameters. We have found, in general that $\NEG{R}_{p}$ contribution
dominates over the SM contribution i.e., by a factor of $10^{2}$ for the
semleptonic decays of $K^{0}$, $10$ for the pure leptonic decays of $K_{L,S}$
, while $10^{2}~$\& $10^{4}$ in case of $B_{s}$ and $B_{d}$ respectively.
This demonstrates the role of $\NEG{R}_{p}$ as a viable model for the study
of NP contribution in rare decays.
\end{abstract}

\section{INTRODUCTION}

Flavor changing neutral currents(FCNC) that mediate different flavored
fermions (quarks) of the same charge are one of the most important tools
searching for physics beyond the Standard Model(SM). This is due to their
rarity owing to GIM mechanism\cite{GIM}. Whereas, FCNC processes involving
leptons are strictly forbidden in SM due to lepton family number
conservation contrary to established experimental facts\cite{T1,T2}, such
processes can only be accommodated through physics beyond the SM, named as
New physics (NP). However, lepton flavor conserving processes can proceed
through both universal and non-universal weak neutral current interactions.
Here universal weak neutral current interactions correspond to the SM
interactions, which are flavor as well as generation blind and Non-universal
weak neutral current interaction represents NP interaction which are flavour
as well as generation sensitive. Analyses of such type of processes are good
for comparative study of different Models. In this paper, we have presented
one class of such type of\ pure leptonic and semileptonic decays of
pseudoscalar mesons involving di-neutrinos in their final state in the frame
work of SM and R-parity violating supersymmetric model.

Leptonic and semileptonic decays of beauty and strange mesons have played an
important role in measuring parameters related to Cabibbo-Kobayashi\newline
-Maskawa (CKM), unitary angles and also in probing CP-violation\cite{BELLE}.
Many New Physics (NP) models like 2HDM \cite{Higgs} and $\NEG{R}_{p}$ MSSM.%
\cite{Rparit} have been explored in these processes\cite{NP} as well. Super
B-factories\cite{SB} also hold a lot of potential in this regard. LHC B also
holds a lot of promise for discovering prospects of new physics in B decays%
\cite{LHCB}.

The Minimal Supersymmetric Standard Model\ ($MSSM$)\  \cite{SUSY}\ is the
most economical version of SUSY. It is also the minimal extension of $SM$%
\cite{SUSY}. MSSM allows processes that violate baryon and lepton number. It
also allows LFV processes. R-parity, a discrete symmetry is imposed to
prevent baryon number, lepton number and flavor violating processes. It is
defined as \newline
$R_{p}=(-1)^{3B+L+2S}$\cite{H1}.\ R-parity conservation is
phenomenologically motivated and if relaxed carefully allows one to analyze
rare and forbidden decays while maintaining the stability of matter\cite{R.
Barbie}. The R-parity violating gauge invariant and renormalizable
superpotential is\cite{H1}%
\begin{equation}
\ W_{\text{\textsl{$\NEG{R}$}}_{\text{\textbf{p}}}}=\frac{1}{2}\lambda
_{ijk}L_{i}L_{j}E_{k}^{c}+\lambda _{ijk}^{^{\prime }}L_{i}Q_{j}D_{k}^{c}+%
\frac{1}{2}\lambda _{ijk}^{^{\prime \prime
}}U_{i}^{^{C}}D_{j}^{^{C}}D_{k}^{^{C}}+\mu _{i}H_{u}L_{i},\  \  \  \  \  \  \  \  \ 
\end{equation}%
where $i,\,j,\,k$\ are generation indices, $L_{i}$\ and $Q_{i}$\ are the
lepton and quark left-handed $SU(2)_{L}$\ doublets and $E^{c}$, $D^{c}$\ are
the charge conjugates of the right-handed leptons and quark singlets,
respectively. $\lambda _{ijk}$, $\lambda _{ijk}^{\prime }$\ and\ $\lambda
_{ijk}^{\prime \prime }$ are Yukawa couplings. The term proportional to $%
\lambda _{ijk}$\ is antisymmetric in first two indices $[i,\,j]$\ and $%
\lambda _{ijk}^{\prime \prime }$ is antisymmetric in\ last$\ $two indices $%
[j,k]$, implying $9(\lambda _{ijk})+27(\lambda _{ijk}^{^{\prime }})+9\left(
\lambda _{ijk}^{^{\prime \prime }}\right) =45$\ independent coupling
constants among which 36 are related to the lepton flavor violation (9 from $%
LLE^{c}$\ and 27 from $LQD^{c}$). We can rotate the last term away without
affecting things of our interest.

In this scenario for detailed illustration we will use the pure and
semileptonic rare decays of pseudoscalar mesons with missing energy i.e. $%
(B^{0}\rightarrow \nu _{\alpha }\overline{\nu }_{\beta },B^{\pm
,0}\rightarrow M^{\pm ,0}\nu _{\alpha }\overline{\nu }_{\beta }$ and $%
B_{X}^{\pm ,0}\rightarrow M^{\pm ,0}\nu _{\alpha }\overline{\nu }_{\beta };$
where $M=\pi ,K$ and subscript $X=S,C).$ At the quark level all $B_{X}^{\pm
,0}\rightarrow M^{\pm ,0}\nu _{\alpha }\overline{\nu }_{\beta }$ decays are
represented by $b\rightarrow q$ $\nu _{\alpha }\overline{\nu }_{\beta }$ $%
(q=d,s),$ and (all these processes can be) divided into two categories on
the bases of lepton flavors i.e.

$1.$ lepton \emph{flavor conserving} $(\alpha =\beta )$ and

$2.$ lepton flavor violating $(\alpha \neq \beta )$ decays.

The \emph{first type} of decays $b\rightarrow q$ $\nu _{\alpha }\overline{%
\nu }_{\alpha }$ ($\alpha =e,\mu ,\tau $) are absent in the SM at tree
level, however are induced by GIM mechanism\cite{GIM} at the quantum loop
level\cite{TL} which makes their effective strength very small, further
suppression caused by the weak mixing angles of the quark flavor rotation
matrix, called Cabibbo-Kobayashi-Maskawa(CKM) matrix \cite{CKM}. These two
suppressions make FCNC decays very rare. Further-more these processes will
provide indirect test of high energy scales through a low energy process.
Such type of processes having only short distance dominant contribution
whereas, long distance contribution is subleading\cite{SOD}. As we are
taking pure and semileptonic decays, which can be accurately predicted in
the standard model (SM) due to the fact that the only relevant hadronic
operators are just the current operators whose matrix elements can be
extracted from their respective leading decays\  \cite{FAR1}.

The second type of decays $b\rightarrow q$ $\nu _{\alpha }\overline{\nu }%
_{\beta }$ ($(\alpha \neq \beta ;$ $\alpha ,\beta =e,\mu ,\tau $) are
strictly forbidden to all orders in the SM due to lepton flavour violation,
so the only possible explanation for these type of processes is Non
Standard/ New interactions. Hence one can say that these are the "golden
channels" for the study of New Physics.

In this paper, we have analyzed above mentioned decays in the SM (first
case) and then in $\NEG{R}_{p}$ violating MSSM. Our focus is to predict the
branching fraction (in some cases) and NP parameters and to develop the
relationship between the parameters of different models. In the forthcoming
section, we will discuss these processes one by one.

\section{s$\rightarrow d\protect \nu _{\protect \alpha }\bar{\protect \nu}_{%
\protect \alpha }~$}

The effective Hamiltonian for the semileptonic($K\rightarrow \pi \nu
_{\alpha }\bar{\nu}_{\alpha },~K\rightarrow \pi ^{0}\nu _{\alpha }\bar{\nu}%
_{\alpha })$ and pure leptonic$~$K$_{L,S}\rightarrow \nu _{\alpha }\bar{\nu}%
_{\alpha }$ processes is given by\cite{Buras}

\begin{equation}
H_{eff}=\sum_{l}C_{SM}(\bar{s}d)_{V-A}(\bar{\nu}_{l}\nu _{l})_{V-A}
\end{equation}

In this case all leptons couple universally with the electroweak gauge
bosons.

where

$\  \  \  \  \  \  \  \  \  \  \  \  \  \  \  \  \  \  \  \  \  \  \  \  \  \  \  \  \  \  \  \ C_{SM}=%
\frac{G_{F}\alpha }{2\sqrt{2}\sin ^{2}\theta _{w}}(V_{cs}^{\ast
}V_{cd}X_{NL}^{l}+V_{ts}^{\ast }V_{td}X(x_{t}))$

and

$\  \  \  \  \  \  \  \  \  \  \  \  \  \  \  \  \  \  \  \  \  \  \  \  \  \  \  \  \  \  \
X(x_{t})=X_{0}(x_{t})+\frac{\alpha _{S}}{4\pi }X_{l}(x_{t})$

and $\  \  \  \  \  \  \  \  \  \  \  \  \  \  \  \  \  \  \  \  \  \  \ x_{t}=\frac{\bar{m}%
_{t}^{2}(\mu _{t})}{M_{W}^{2}},$\ $\mu _{t}=O(m_{t}).$

In MSSM the relevant effective Lagrangian for the decay process $%
K\rightarrow \pi \nu _{\alpha }\bar{\nu}_{\alpha }~$is given by\cite{FAR1}

\begin{equation}
L_{\QTR{sl}{\NEG{R}\;}_{\text{\textbf{p}}}}^{eff}\left( s\longrightarrow
d+\nu _{\alpha }+\bar{\nu}_{\alpha }\right) =\frac{4G_{F}}{\sqrt{2}}\left[ 
\begin{array}{c}
A_{\alpha \beta }^{sd}\left( \overline{\nu }_{\alpha }\gamma ^{\mu }P_{L}\nu
_{\alpha }\right) \left( \overline{d}\gamma _{\mu }P_{R}s\right)%
\end{array}%
\right] .
\end{equation}%
Where $\alpha =e,\mu .\ $The first term in eq. (2) comes from the down
squark exchange (where $d$ and $s$\ are down type quarks). The dimensionless
coupling constant $A_{\alpha \alpha }^{sd}\ $is given by

\begin{equation}
A_{\alpha \alpha }^{sd}=\frac{\sqrt{2}}{4G_{F}}\underset{k=1}{\overset{3}{%
\sum }}\frac{\lambda _{\alpha k1}^{\prime }\lambda _{\alpha k2}^{\prime \ast
}}{2m_{\widetilde{d_{k}^{c}}}^{2}},
\end{equation}

The differential decay rate\ for semileptonic decay processes is given by

\begin{equation}
\frac{d\Gamma }{dq^{2}}=\frac{1}{2^{5}\pi ^{5}}\lambda
^{3/2}(1,r,s)m_{K}^{3}\left \vert f_{p}^{+}(q^{2})\right \vert ^{2}\left
\vert C_{l}\right \vert
\end{equation}

where, $C_{l}=C_{SM}+C_{\NEG{R}_{p}}$ with $C_{\NEG{R}_{p}}=\frac{\lambda
_{\alpha k1}^{\prime }\lambda _{\alpha k2}^{\prime \ast }}{m_{\widetilde{%
d_{k}^{c}}}^{2}}$

The decay rate\ for pure leptonic decay processes is given by

\begin{equation}
\Gamma (s\rightarrow d\nu _{l}\bar{\nu}_{l})=\frac{1}{8\pi }m_{K}^{3}\left
\vert f_{p}(q^{2})\right \vert ^{2}\left \vert \frac{2m_{l}}{m_{K}}%
C_{l}\right \vert ^{2}
\end{equation}

\section{$b\rightarrow d(s)\protect \nu _{\protect \alpha }\protect \nu _{%
\protect \alpha }$}

In MSSM, the relevant effective Lagrangian for the decay process $%
B\rightarrow \pi (K)\nu _{\alpha }\bar{\nu}_{\alpha }~$is given by\cite{FAR1}

\begin{equation}
L_{\QTR{sl}{\NEG{R}\;}_{\text{\textbf{p}}}}^{eff}\left( b\longrightarrow
d(s)+\nu _{\alpha }+\bar{\nu}_{\alpha }\right) =\frac{4G_{F}}{\sqrt{2}}\left[
\begin{array}{c}
A_{\alpha \alpha }^{bd(s)}\left( \overline{\nu }_{\alpha }\gamma ^{\mu
}P_{L}\nu _{\alpha }\right) \left( \bar{b}\gamma _{\mu }P_{R}d(s)\right)%
\end{array}%
\right] .
\end{equation}%
Where $\alpha =e,\mu .\ $The first term in eq. (2) comes from the down
squark exchange (where $b$ and $d(s)$\ are down type quarks). The
dimensionless coupling constant $A_{\alpha \alpha }^{bd(s)}\ $is given by

\begin{equation}
A_{\alpha \alpha }^{bd(s)}=\frac{\sqrt{2}}{4G_{F}}\underset{k=1}{\overset{3}{%
\sum }}\frac{\lambda _{\alpha k1(2)}^{\prime }\lambda _{\alpha k3}^{\prime
\ast }}{2m_{\widetilde{d_{k}^{c}}}^{2}},
\end{equation}

The differential decay rate\ for semileptonic decay processes is given by

\begin{equation}
\frac{d\Gamma }{dq^{2}}=\frac{1}{2^{5}\pi ^{5}}\lambda
^{3/2}(1,r,s)m_{B}^{3}\left \vert f_{p}^{+}(q^{2})\right \vert ^{2}\left
\vert C_{l}\right \vert
\end{equation}

where\medskip

\begin{center}
$C_{l}=C_{SM}+C_{\NEG{R}_{p}}$
\end{center}

with\medskip

\begin{center}
$C_{SM}=\frac{G_{F}\alpha }{2\sqrt{2}\sin ^{2}\theta _{w}}(V_{cb}^{\ast
}V_{cd(s)}X_{NL}^{l}+V_{tb}^{\ast }V_{td(s)}X(x_{t}))$\medskip
\end{center}

with\medskip

\begin{center}
$X(x_{t})=X_{0}(x_{t})+\frac{\alpha _{S}}{4\pi }X_{l}(x_{t})$ and $x_{t}=%
\frac{\bar{m}_{t}^{2}(\mu _{t})}{M_{W}^{2}},$\ $\mu _{t}=O(m_{t}).$\medskip
\end{center}

and\medskip

\begin{center}
$C_{\NEG{R}_{p}}=\frac{\lambda _{\alpha k1(2)}^{\prime }\lambda _{\alpha
k3}^{\prime \ast }}{m_{\widetilde{d_{k}^{c}}}^{2}}$
\end{center}

The decay rate\ for pure leptonic decay processes is given by

\begin{equation}
\Gamma (b\rightarrow d(s)\nu _{l}\bar{\nu}_{l})=\frac{1}{8\pi }%
m_{B}^{3}\left \vert f_{p}(q^{2})\right \vert ^{2}\left \vert \frac{2m_{l}}{%
m_{B}}C_{l}\right \vert ^{2}
\end{equation}

\section{Results And Discussions}

We have carried out analysis of hypercharge changing two and three body
decay processes of pseudoscalar mesons~($M\rightarrow X\nu \bar{\nu}%
;~M\rightarrow \nu \bar{\nu}$) where$~M=K,B$ and$~X=\pi ,K$. The feynman
diagrams and table listing experimental data related to these processes are
given in Fig. (1) and Table [I] respectively

%TCIMACRO{%
%\TeXButton{Feynman Diagrams}{\begin{figure}[H]
%\begin{center}
%\subfigure[] {\includegraphics[scale=0.3]{1.png}}
%\subfigure[] {\includegraphics[scale=0.3]{2.png}}
%\subfigure[] {\includegraphics[scale=0.3]{3.png}}
%\caption{Feynman diagrams of  (a) $b \rightarrow s \nu _{\alpha} \overline{\nu} _{\alpha} (b)  b \rightarrow d \nu _{\alpha} \overline{\nu} _{\alpha} (c) s \rightarrow d \nu _{\alpha} \overline{\nu} _{\alpha}$}
%\end{center}
%\end{figure}}}%
%BeginExpansion
\begin{figure}[H]
\begin{center}
\subfigure[] {\includegraphics[scale=0.3]{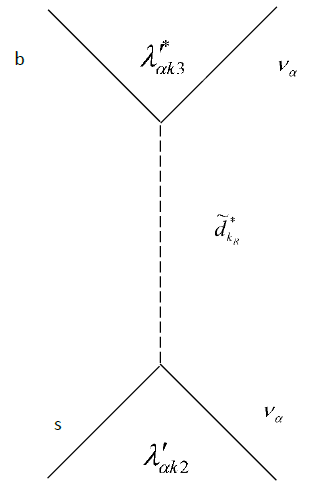}}
\subfigure[] {\includegraphics[scale=0.3]{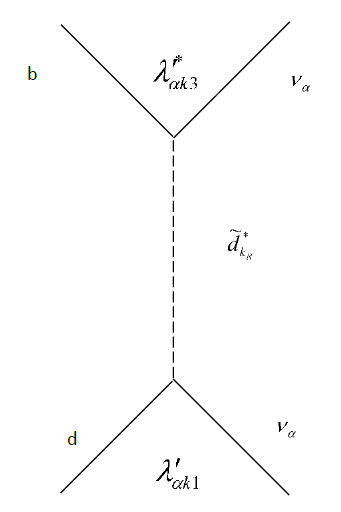}}
\subfigure[] {\includegraphics[scale=0.3]{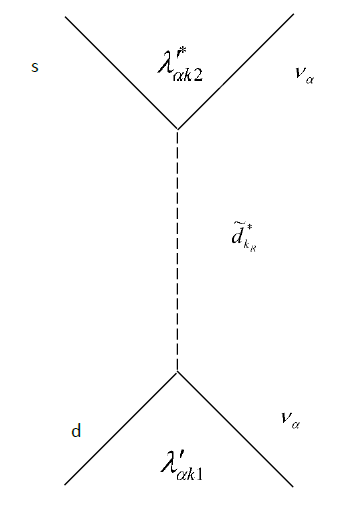}}
\caption{Feynman diagrams of  (a) $b \rightarrow s \nu _{\alpha} \overline{\nu} _{\alpha} (b)  b \rightarrow d \nu _{\alpha} \overline{\nu} _{\alpha} (c) s \rightarrow d \nu _{\alpha} \overline{\nu} _{\alpha}$}
\end{center}
\end{figure}%
%EndExpansion

%TCIMACRO{%
%\TeXButton{Main Table }{\begin{table}[H]
%\begin{center}
%{\includegraphics[scale=0.5]{Table1.jpg}}
%\caption{Table listing the property of processes under discussion}
%\end{center}
%\end{table}}}%
%BeginExpansion
\begin{table}[H]
\begin{center}
{\includegraphics[scale=0.5]{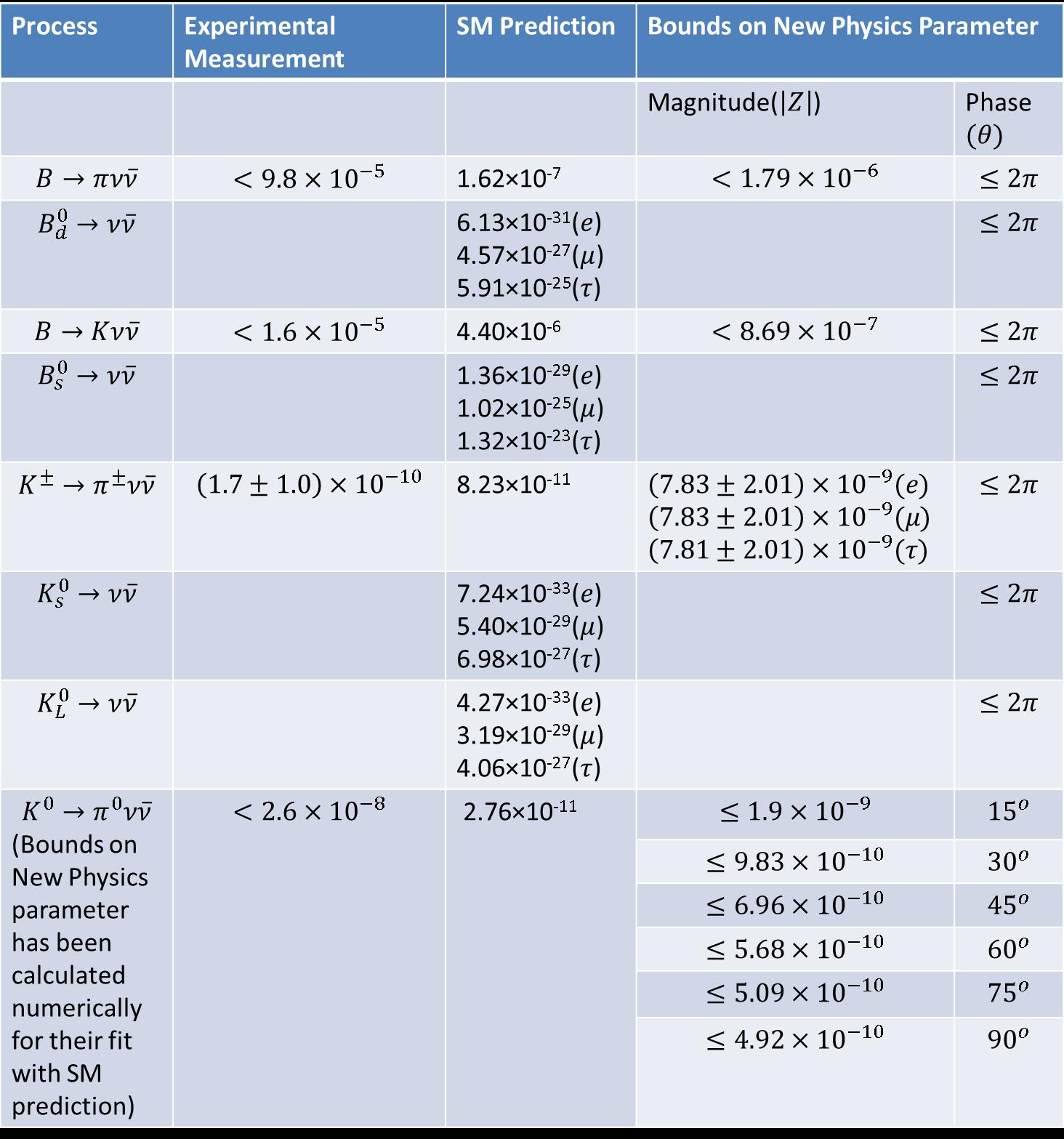}}
\caption{Table listing the property of processes under discussion}
\end{center}
\end{table}%
%EndExpansion

First, we will discuss the results related to semileptonic decay processes
followed by pure leptonic decays. We have plotted graphs in Figs. (2 and 3)
for the study of process $K\rightarrow \pi \nu _{\alpha }\bar{\nu}_{a}.$
These plots relate the branching fraction of the said process with the
magnitude and phase of New Physics(\textbf{NP}) parameters($\left \vert
z(\lambda _{ik1}^{\prime }~\lambda _{ik2}^{\prime \ast })\right \vert $ and $%
\theta $). Plots in Fig. (2) represent the allowed region for NP parameters
for a specific value of branching fraction at $\pm 1\sigma $ level. All four
plots (comprising of \textit{Unpolarized}(a) and \textit{Polarized}(b-d))
show that the maximum magnitude of NP parameter oscillates w.r.t. its phase
in general. The plot in Fig. (2a) shows a particular pattern at given error
level at $-1\sigma $ level of measured branching fraction($0.7\times 10^{-7}$%
). It clearly shows that only a narrow range of phase of NP parameter($%
\left
\vert \theta \right \vert \leq \frac{\pi }{4}$) is allowed for given $%
-1\sigma $ level. The bounds on the magnitude of NP parameter are given in
Table [2]. Tables [2a and 2b] show the same pattern as observed in Figs. (2a
and 3a) numerically. Since Yukawa couplings for R-parity violation are
identical for the processes $(K\rightarrow \pi \nu _{\alpha }\bar{\nu}%
_{\alpha },~K^{0}\rightarrow \pi ^{0}\nu _{\alpha }\bar{\nu}_{\alpha
},~K_{L,S}\rightarrow \nu _{\alpha }\bar{\nu}_{\alpha })$, the maximum
limits for $K\rightarrow \pi \nu _{\alpha }\bar{\nu}_{\alpha }$ are used for
calculating \textit{NP }contribution to branching fraction of other
processes.

Fig. (4) shows the relationship between\ the branching fraction of the
process ($K^{0}\rightarrow \pi ^{0}\nu _{\alpha }\bar{\nu}_{a}$), magnitude
and phase of New Physics(\textbf{NP}) parameters\newline
($\left \vert z(\lambda _{ik1}^{\prime }~\lambda _{ik2}^{\prime \ast
})\right \vert $ and $\theta $). These plots comprise of the allowed region
for NP parameters for a specific value of branching fraction at $\pm 1\sigma 
$ level. All four plots (comprising of \textit{Unpolarized}(a) and \textit{%
Polarized}(b-d)) show that the maximum magnitude of NP parameter follows a 
\textit{catenary}(hanging chain) pattern, with the bottom level smoothening
with decreasing error levels. The bounds on the magnitude of NP parameter
and possible NP contribution are in Table [3], which shows that $\NEG{R}_{p}$
MSSM enhances SM contribution by order of magnitude \symbol{126}$10^{2}$ but
an order of \symbol{126}$10^{2}$ below experimental limits.

Figs.\ (5 and 6) relate the branching fraction of the process ($B\rightarrow
\pi \nu _{\alpha }\bar{\nu}_{a}).$with the magnitude and phase of \textbf{NP}
parameters($\left \vert z(\lambda _{ik1}^{\prime }~\lambda _{ik2}^{\prime
\ast })\right \vert $ and $\theta $). Graphs in Fig. (5) illustrate the
allowed region for NP parameters for a specific value of branching fraction
at $\pm 1\sigma $ level. All four plots (comprising of \textit{Unpolarized}%
(a) and \textit{Polarized}(b-d)) show that the maximum magnitude of NP
parameter oscillates gently w.r.t. its phase in general. Similarly, plots in
Fig. (6) represent the variation of branching fraction w.r.t. the magnitude
of NP parameter($\left \vert z(\lambda _{ik1}^{\prime }~\lambda
_{ik2}^{\prime \ast })\right \vert $). All four plots demonstrates the
gentle oscillation behavior as observed in Fig. (5) with sharply distinct
curves for different values of phases of NP parameter($\theta $). The bounds
on the magnitude of NP parameter are given in Table [6].

Plots in Fig. (7) describe the branching fraction of the process ($%
B\rightarrow K\nu _{\alpha }\bar{\nu}_{a})$ with the magnitude and phase of 
\textbf{NP} parameters($\left \vert z(\lambda _{ik3}^{\prime }~\lambda
_{ik2}^{\prime \ast })\right \vert $ and $\theta $). Graphs in Fig. (7)
depict the allowed region for NP parameters for a specific value of
branching fraction bounded by the experimental limit, while plots in Fig.
(3) represent the variation of branching fraction w.r.t. the magnitude of NP
parameter($\left \vert z(\lambda _{ik3}^{\prime }~\lambda _{ik2}^{\prime \ast
})\right \vert $). All four plots (comprising of \textit{Unpolarized}(a) and 
\textit{Polarized}(b-d)) show that the maximum magnitude of NP parameter
oscillates w.r.t. its phase in general. The plot in Fig. (2a) shows a
particular pattern below given limiting branching fraction ($\leq 4\times
10^{-6}$). This particular pattern shows that only a narrow range of phase($%
\theta $) of NP parameter is allowed in that case.

For pure leptonic decays of strange mesons involving neutrinos($%
K_{L,S}\rightarrow \nu _{\alpha }\bar{\nu}_{\alpha }$), there is no
experimental data available. Therefore, we use limits derived from $%
(K\rightarrow \pi \nu _{\alpha }\bar{\nu}_{\alpha })$ to calculate NP
contribution to these processes. The bounds on the magnitude of NP parameter
and possible NP contribution are given in Tables [4 and 5] for the decay of $%
K_{S,L}$ respectively, which shows that $\NEG{R}_{p}$ MSSM enhances SM
contribution by order $\symbol{126}10$ for $K_{S,L}.$

For pure leptonic decays of beauty involving neutrinos($B_{d,s}\rightarrow
\nu _{\alpha }\bar{\nu}_{\alpha }$), there is no experimental data available
for these processes, we use limits derived from $(B\rightarrow (\pi ,K)\nu
_{\alpha }\bar{\nu}_{\alpha })$ to calculate NP contributions to these
processes. \ The bounds on the magnitude of NP parameter and possible NP
contribution are given in Tables [8 and 9] for the decay of $B_{d,s}$
respectively, which shows that $\NEG{R}_{p}$ MSSM enhances SM contribution
by order of magnitude $\symbol{126}10$ for $B_{s}$ and $10^{4}$ for $B_{d}.$

\newpage

%TCIMACRO{%
%\TeXButton{Bound}{\begin{table}[H]
%\subfigure[] {\includegraphics[scale=0.3]{Picture1.png}}
%\subfigure[] {\includegraphics[scale=0.3]{Picture2.png}}
%\subfigure[] {\includegraphics[scale=0.3]{Picture3.png}}
%\subfigure[] {\includegraphics[scale=0.3]{Picture4.png}}
%\subfigure[] {\includegraphics[scale=0.3]{Picture5.png}}
%\caption{Bounds on NP Parameter ($\left|z(\lambda '_{i\text{jk}} \lambda '_{l\text{mn}}\right)|$,$\theta)$ for $K \rightarrow \pi \nu _{\alpha} \overline{\nu} _{\alpha},  \alpha$ ($Br_{SM}$)  are  (a) UnPolarized (8.63$\times10^{-11})$ (b) e (2.89$\times10^{-11})$(c)$ \mu$(2.89$\times10^{-11})$(d)$ \tau$ (2.85$\times10^{-11})$. Experimental limits are $((1.7\pm1.0)\times10^{-10})$}
%\end{table}}}%
%BeginExpansion
\begin{table}[H]
\subfigure[] {\includegraphics[scale=0.3]{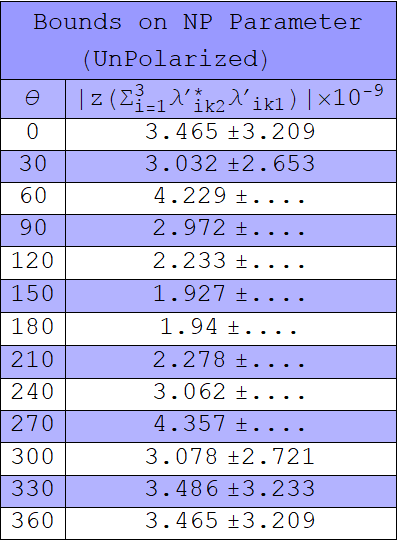}}
\subfigure[] {\includegraphics[scale=0.3]{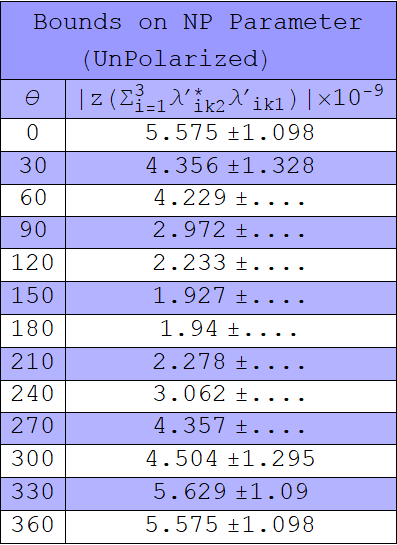}}
\subfigure[] {\includegraphics[scale=0.3]{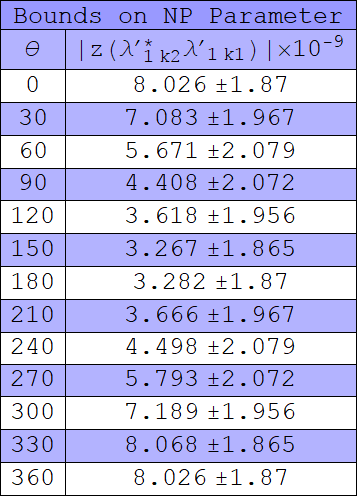}}
\subfigure[] {\includegraphics[scale=0.3]{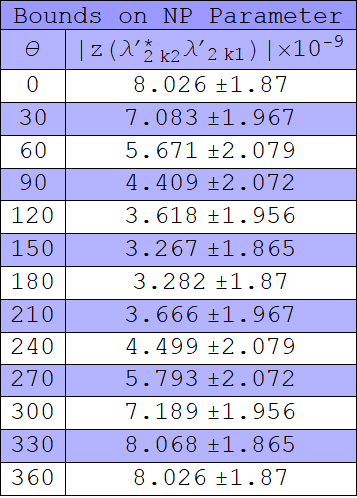}}
\subfigure[] {\includegraphics[scale=0.3]{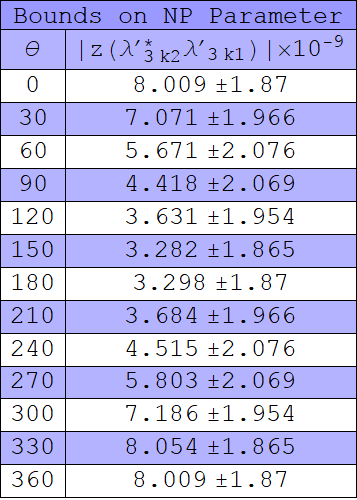}}
\caption{Bounds on NP Parameter ($\left|z(\lambda '_{i\text{jk}} \lambda '_{l\text{mn}}\right)|$,$\theta)$ for $K \rightarrow \pi \nu _{\alpha} \overline{\nu} _{\alpha},  \alpha$ ($Br_{SM}$)  are  (a) UnPolarized (8.63$\times10^{-11})$ (b) e (2.89$\times10^{-11})$(c)$ \mu$(2.89$\times10^{-11})$(d)$ \tau$ (2.85$\times10^{-11})$. Experimental limits are $((1.7\pm1.0)\times10^{-10})$}
\end{table}%
%EndExpansion

%TCIMACRO{%
%\TeXButton{Contour Plot 1}{\begin{figure}[H]
%\begin{center}
%\subfigure[] {\includegraphics[scale=0.3]{A1.png}}
%\subfigure[] {\includegraphics[scale=0.3]{A2.png}}
%\subfigure[] {\includegraphics[scale=0.3]{A3.png}}
%\subfigure[] {\includegraphics[scale=0.3]{A4.png}}
%\end{center}
%\caption{Allowed region of general New Physics(NP) Parameter $ (z,\theta)$ for $K \rightarrow \pi \nu _{\alpha} \overline{\nu} _{\alpha},  \alpha$  are  (a) UnPolarized (b) e (c) $\mu (d) \tau$. The three contours belong to branching fraction at $[0.7,1.7,2.7]$}
%\end{figure}}}%
%BeginExpansion
\begin{figure}[H]
\begin{center}
\subfigure[] {\includegraphics[scale=0.3]{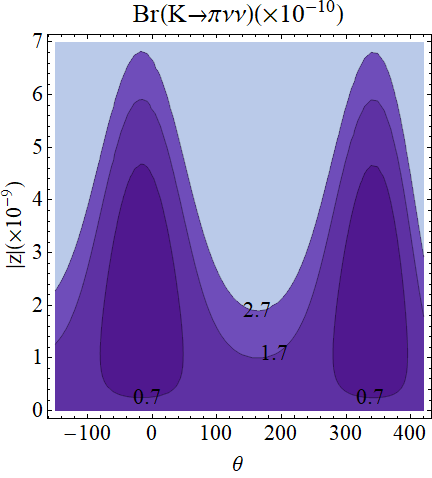}}
\subfigure[] {\includegraphics[scale=0.3]{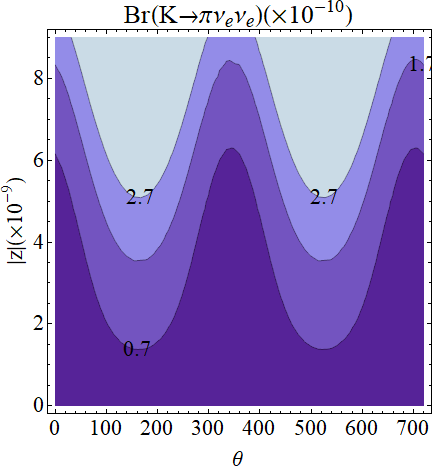}}
\subfigure[] {\includegraphics[scale=0.3]{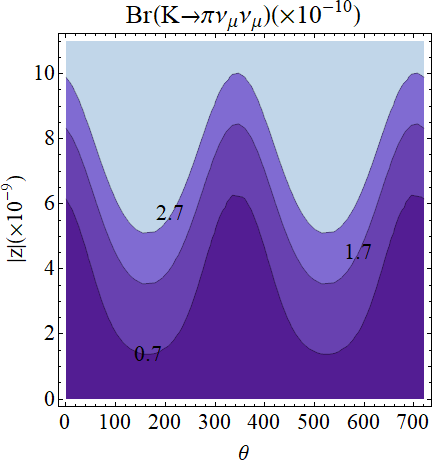}}
\subfigure[] {\includegraphics[scale=0.3]{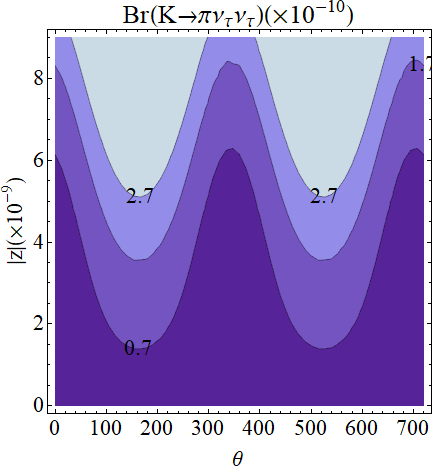}}
\end{center}
\caption{Allowed region of general New Physics(NP) Parameter $ (z,\theta)$ for $K \rightarrow \pi \nu _{\alpha} \overline{\nu} _{\alpha},  \alpha$  are  (a) UnPolarized (b) e (c) $\mu (d) \tau$. The three contours belong to branching fraction at $[0.7,1.7,2.7]$}
\end{figure}%
%EndExpansion

%TCIMACRO{%
%\TeXButton{Linear Plot1}{\begin{figure}[H]
%\begin{center}
%\subfigure[] {\includegraphics[scale=0.29]{Graph1.png}}
%\subfigure[] {\includegraphics[scale=0.29]{Graph2.png}}
%\subfigure[] {\includegraphics[scale=0.29]{Graph3.png}}
%\subfigure[] {\includegraphics[scale=0.29]{Graph4.png}}
%\end{center}
%\caption{Variations of branching fraction w.r.t NP Parameter ($\left|z(\lambda '_{i\text{jk}} \lambda '_{l\text{mn}}\right)|$,$\theta)$ for $K \rightarrow \pi \nu _{\alpha} \overline{\nu} _{\alpha},  \alpha$  are  (a) UnPolarized (b) e (c) $\mu (d) \tau$. The three bounds belong to branching fraction at $[0.7,1.7,2.7]$}
%\end{figure}}}%
%BeginExpansion
\begin{figure}[H]
\begin{center}
\subfigure[] {\includegraphics[scale=0.29]{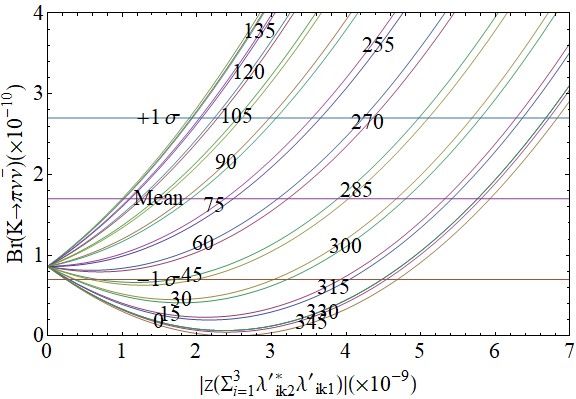}}
\subfigure[] {\includegraphics[scale=0.29]{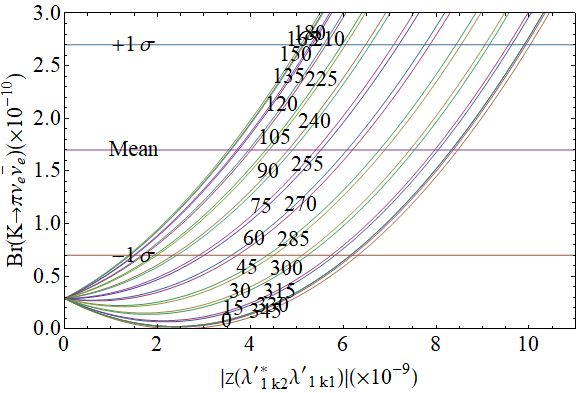}}
\subfigure[] {\includegraphics[scale=0.29]{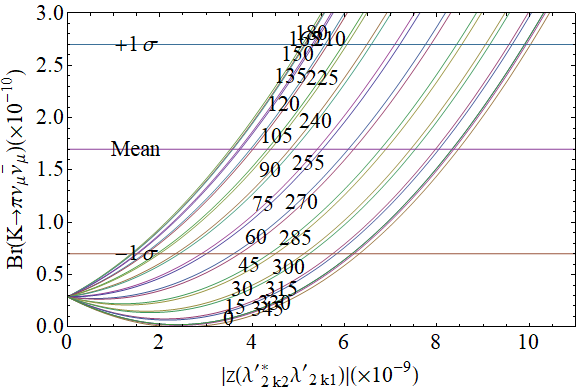}}
\subfigure[] {\includegraphics[scale=0.29]{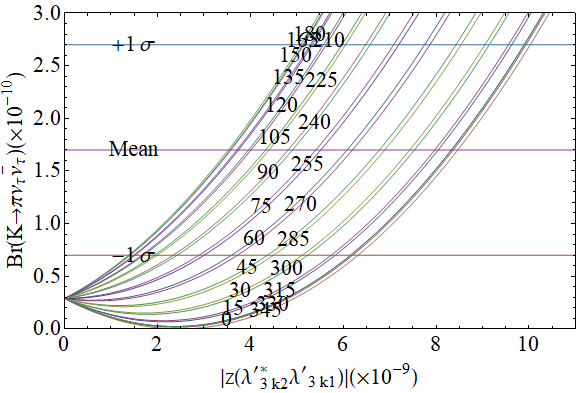}}
\end{center}
\caption{Variations of branching fraction w.r.t NP Parameter ($\left|z(\lambda '_{i\text{jk}} \lambda '_{l\text{mn}}\right)|$,$\theta)$ for $K \rightarrow \pi \nu _{\alpha} \overline{\nu} _{\alpha},  \alpha$  are  (a) UnPolarized (b) e (c) $\mu (d) \tau$. The three bounds belong to branching fraction at $[0.7,1.7,2.7]$}
\end{figure}%
%EndExpansion

%TCIMACRO{%
%\TeXButton{Bound Table 1}{\begin{table}[H]
%\subfigure[] {\includegraphics[scale=0.22]{Picture6.png}}
%\subfigure[] {\includegraphics[scale=0.22]{Picture61.png}}
%\subfigure[] {\includegraphics[scale=0.22]{Picture62.png}}
%\subfigure[] {\includegraphics[scale=0.22]{Picture63.png}}
%\caption{ Bounds on NP Parameter (derived from $K \rightarrow \pi \nu _{\alpha} \overline{\nu} _{\alpha},  \alpha$)  ($\left|\lambda '_{i\text{jk}} \lambda '_{l\text{mn}}\right|$,$\theta)$ for $K^0 \rightarrow \pi^0 \nu _{\alpha} \overline{\nu} _{\alpha},  \alpha$ ($Br_{SM}$)  are  (a) UnPolarized  (2.94.$\times10^{-11})$ and (9.78.$\times10^{-12})$ for (b) e (c) $\mu (d) \tau$. Experimental bound on the process is (2.6 $\times10^{-8})$}
%\end{table}}}%
%BeginExpansion
\begin{table}[H]
\subfigure[] {\includegraphics[scale=0.22]{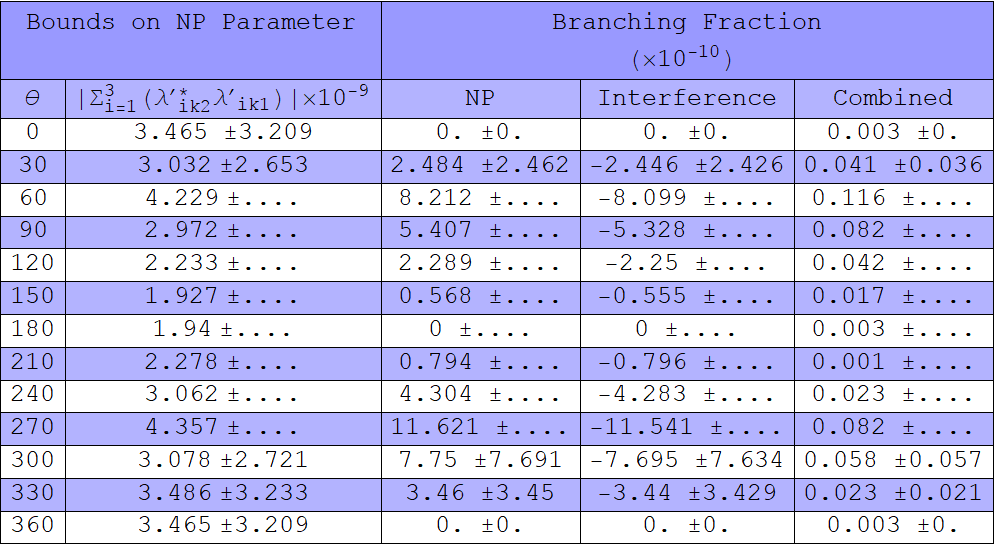}}
\subfigure[] {\includegraphics[scale=0.22]{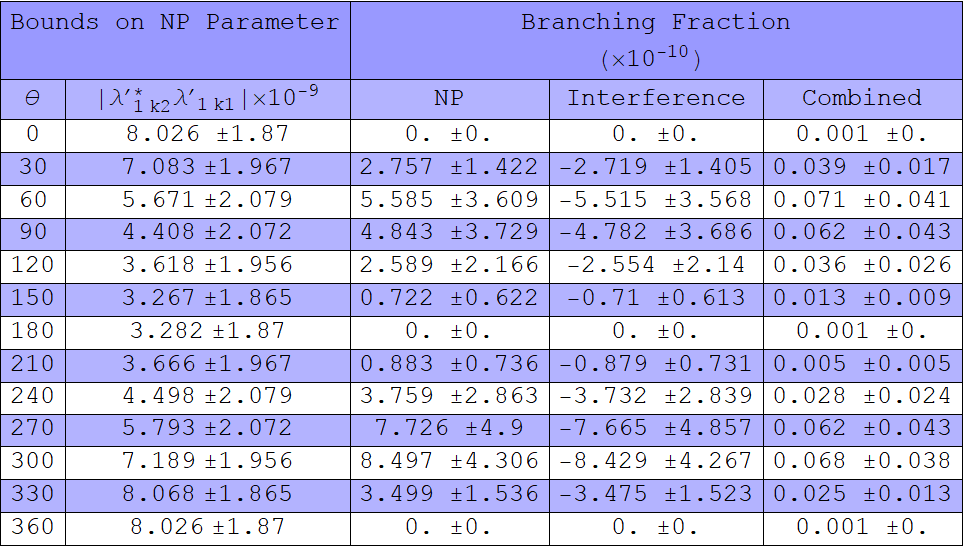}}
\subfigure[] {\includegraphics[scale=0.22]{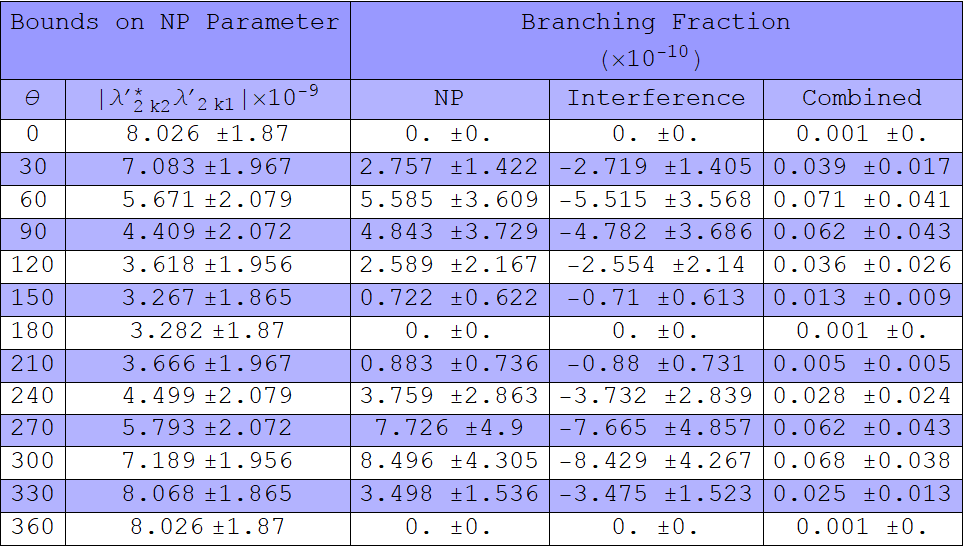}}
\subfigure[] {\includegraphics[scale=0.22]{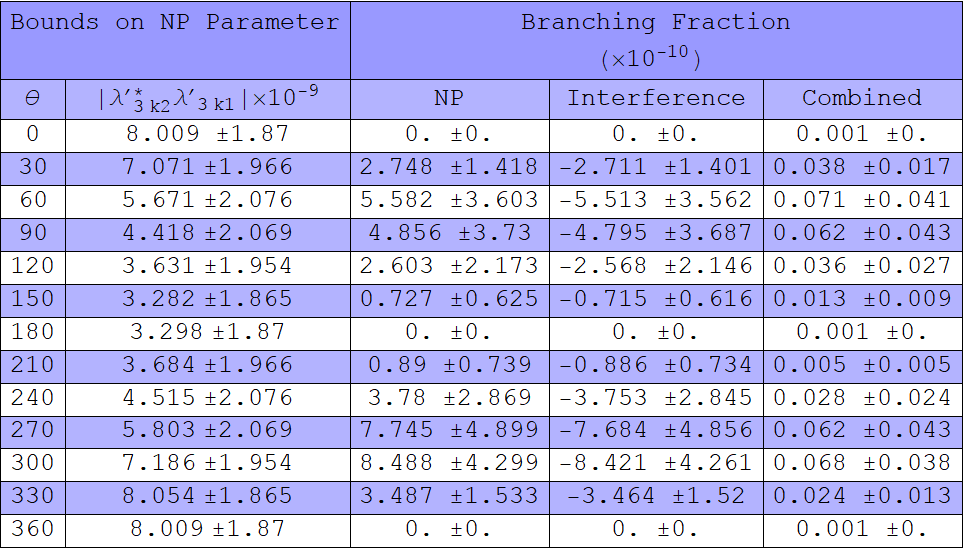}}
\caption{ Bounds on NP Parameter (derived from $K \rightarrow \pi \nu _{\alpha} \overline{\nu} _{\alpha},  \alpha$)  ($\left|\lambda '_{i\text{jk}} \lambda '_{l\text{mn}}\right|$,$\theta)$ for $K^0 \rightarrow \pi^0 \nu _{\alpha} \overline{\nu} _{\alpha},  \alpha$ ($Br_{SM}$)  are  (a) UnPolarized  (2.94.$\times10^{-11})$ and (9.78.$\times10^{-12})$ for (b) e (c) $\mu (d) \tau$. Experimental bound on the process is (2.6 $\times10^{-8})$}
\end{table}%
%EndExpansion

%TCIMACRO{%
%\TeXButton{Contour Plot 2}{\begin{figure}[H]
%\begin{center}
%\subfigure[] {\includegraphics[scale=0.26]{B1.png}}
%\subfigure[] {\includegraphics[scale=0.26]{B2.png}}
%\subfigure[] {\includegraphics[scale=0.26]{B3.png}}
%\subfigure[] {\includegraphics[scale=0.26]{B4.png}}
%\end{center}
%\caption{Allowed regions of general New Physics(NP) Parameter $ (z,\theta)$ for $K^0 \rightarrow \pi^0 \nu _{\alpha} \overline{\nu} _{\alpha},  \alpha$  are (a) UnPolarized (b) e (c) $\mu (d) \tau$. The three contours belong to branching fraction at $[0.3,1.3,2.6]$}
%\end{figure}}}%
%BeginExpansion
\begin{figure}[H]
\begin{center}
\subfigure[] {\includegraphics[scale=0.26]{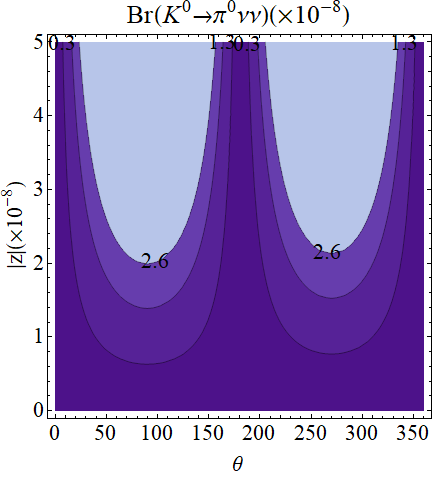}}
\subfigure[] {\includegraphics[scale=0.26]{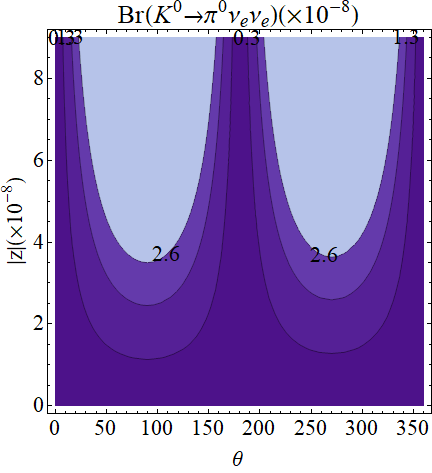}}
\subfigure[] {\includegraphics[scale=0.26]{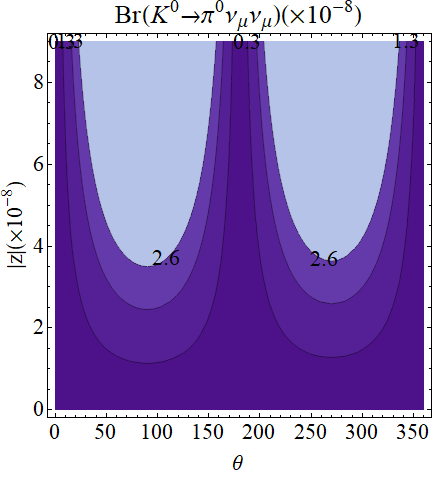}}
\subfigure[] {\includegraphics[scale=0.26]{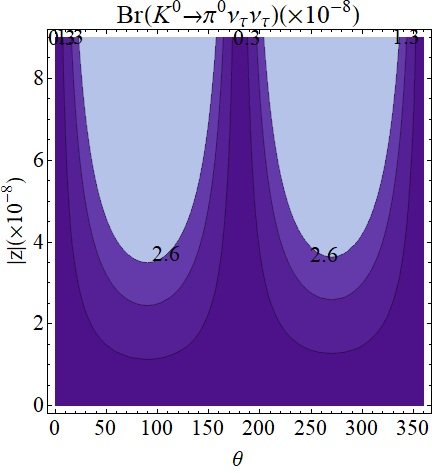}}
\end{center}
\caption{Allowed regions of general New Physics(NP) Parameter $ (z,\theta)$ for $K^0 \rightarrow \pi^0 \nu _{\alpha} \overline{\nu} _{\alpha},  \alpha$  are (a) UnPolarized (b) e (c) $\mu (d) \tau$. The three contours belong to branching fraction at $[0.3,1.3,2.6]$}
\end{figure}%
%EndExpansion

%TCIMACRO{%
%\TeXButton{Linear Plot2}{\begin{figure}[H]
%\begin{center}
%\subfigure[] {\includegraphics[scale=0.28]{NPkpi1.png}}
%\subfigure[] {\includegraphics[scale=0.28]{NPkpi2.png}}
%\subfigure[] {\includegraphics[scale=0.28]{NPkpi3.png}}
%\subfigure[] {\includegraphics[scale=0.28]{NPkpi4.png}}
%\end{center}
%\caption{Variations of NP contribution w.r.t.NP Parameter ($\left|\lambda '_{i\text{jk}} \lambda '_{l\text{mn}}\right|$,$\theta)$ for $K^0 \rightarrow \pi^0 \nu _{\alpha} \overline{\nu} _{\alpha},  \alpha$  is  (a) UnPolarized (b) e (c) $\mu (d) \tau$.}
%\end{figure}}}%
%BeginExpansion
\begin{figure}[H]
\begin{center}
\subfigure[] {\includegraphics[scale=0.28]{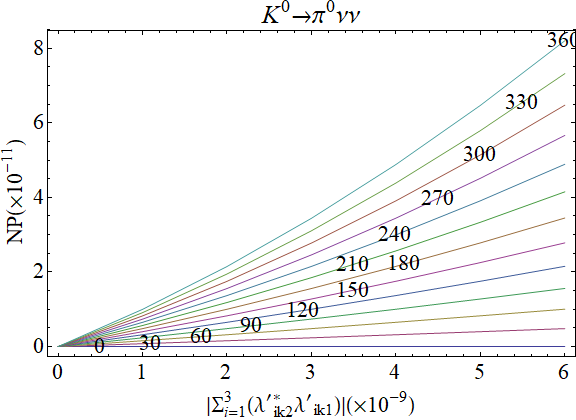}}
\subfigure[] {\includegraphics[scale=0.28]{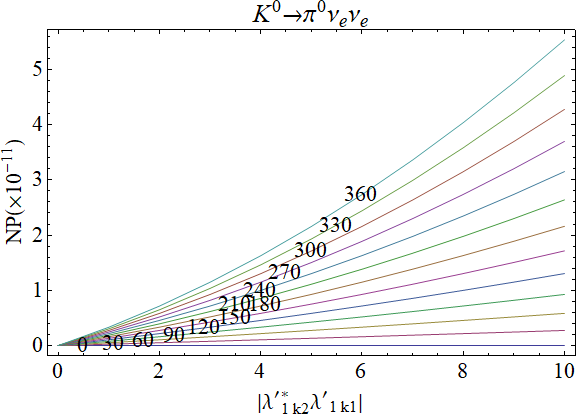}}
\subfigure[] {\includegraphics[scale=0.28]{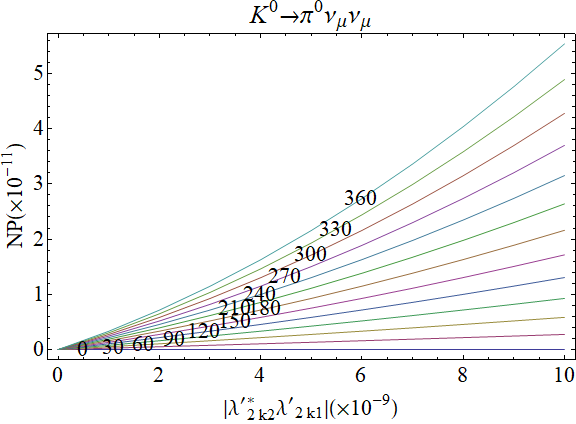}}
\subfigure[] {\includegraphics[scale=0.28]{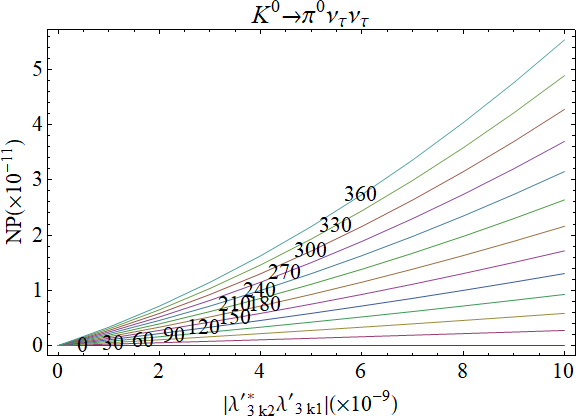}}
\end{center}
\caption{Variations of NP contribution w.r.t.NP Parameter ($\left|\lambda '_{i\text{jk}} \lambda '_{l\text{mn}}\right|$,$\theta)$ for $K^0 \rightarrow \pi^0 \nu _{\alpha} \overline{\nu} _{\alpha},  \alpha$  is  (a) UnPolarized (b) e (c) $\mu (d) \tau$.}
\end{figure}%
%EndExpansion

%TCIMACRO{%
%\TeXButton{Bound Table 2}{\begin{table}[H]
%\subfigure[] {\includegraphics[scale=0.25]{Picture7.png}}
%\subfigure[] {\includegraphics[scale=0.25]{Picture8.png}}
%\subfigure[] {\includegraphics[scale=0.25]{Picture9.png}}
%\subfigure[] {\includegraphics[scale=0.25]{Picture10.png}}
%\caption{ Bounds on NP Parameter ($\left|z(\lambda '_{i\text{jk}} \lambda '_{l\text{mn}}\right)|$,$\theta)$ for $B \rightarrow \pi \nu _{\alpha} \overline{\nu} _{\alpha},  \alpha$ ($Br_{SM}$)  are  (a) UnPolarized (1.73$\times10^{-7})$ and (5.76$\times10^{-8})$ for (b) e (c) $\mu (d) \tau$. Experimental bound on the process is (9.8$\times10^{-5}$). }
%\end{table}}}%
%BeginExpansion
\begin{table}[H]
\subfigure[] {\includegraphics[scale=0.25]{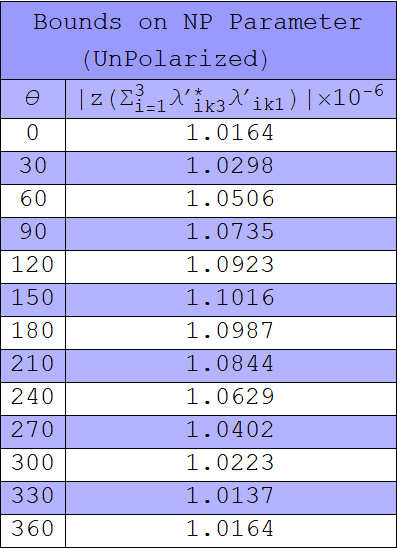}}
\subfigure[] {\includegraphics[scale=0.25]{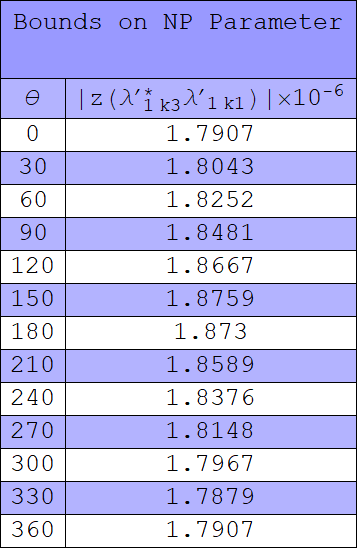}}
\subfigure[] {\includegraphics[scale=0.25]{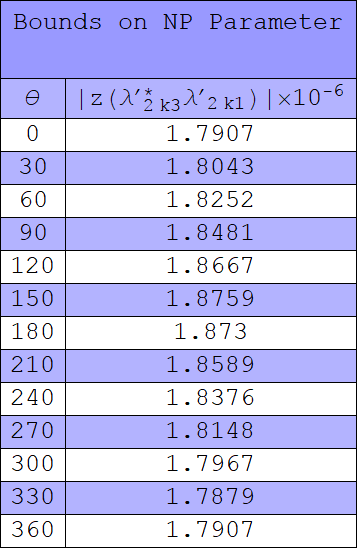}}
\subfigure[] {\includegraphics[scale=0.25]{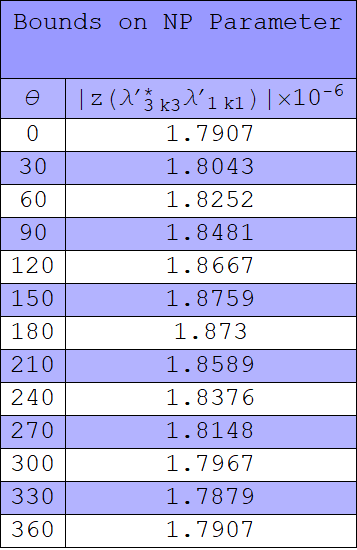}}
\caption{ Bounds on NP Parameter ($\left|z(\lambda '_{i\text{jk}} \lambda '_{l\text{mn}}\right)|$,$\theta)$ for $B \rightarrow \pi \nu _{\alpha} \overline{\nu} _{\alpha},  \alpha$ ($Br_{SM}$)  are  (a) UnPolarized (1.73$\times10^{-7})$ and (5.76$\times10^{-8})$ for (b) e (c) $\mu (d) \tau$. Experimental bound on the process is (9.8$\times10^{-5}$). }
\end{table}%
%EndExpansion

%TCIMACRO{%
%\TeXButton{Contour Plot 3}{\begin{figure}[H]
%\begin{center}
%\subfigure[] {\includegraphics[scale=0.26]{C1.png}}
%\subfigure[] {\includegraphics[scale=0.26]{C2.png}}
%\subfigure[] {\includegraphics[scale=0.26]{C3.png}}
%\subfigure[] {\includegraphics[scale=0.26]{C4.png}}
%\end{center}
%\caption{Allowed regions of general New Physics(NP) Parameter $ (z,\theta)$ for $B \rightarrow \pi \nu _{\alpha} \overline{\nu} _{\alpha},  \alpha$  are  (a) UnPolarized (b) e (c) $\mu (d) \tau$. The three contours belong to branching fraction at $[1.4-9.8]$}
%\end{figure}}}%
%BeginExpansion
\begin{figure}[H]
\begin{center}
\subfigure[] {\includegraphics[scale=0.26]{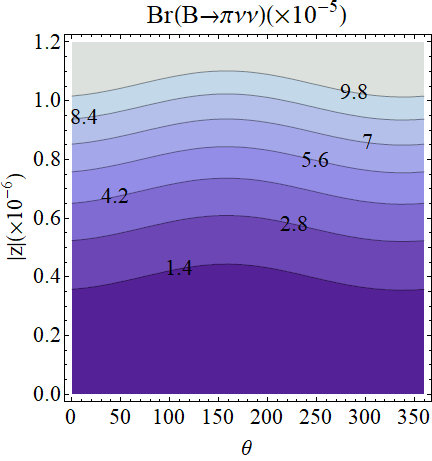}}
\subfigure[] {\includegraphics[scale=0.26]{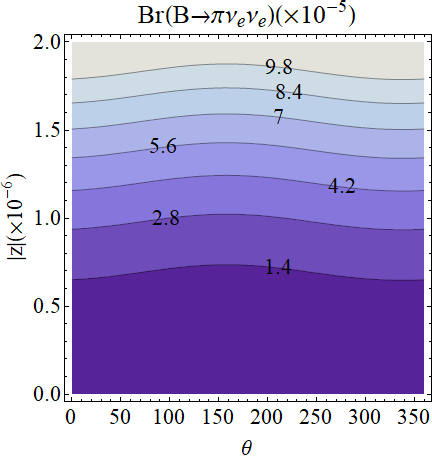}}
\subfigure[] {\includegraphics[scale=0.26]{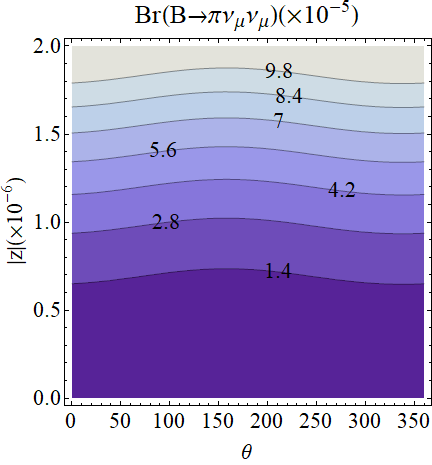}}
\subfigure[] {\includegraphics[scale=0.26]{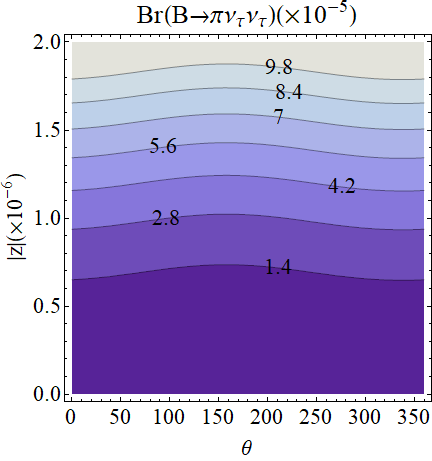}}
\end{center}
\caption{Allowed regions of general New Physics(NP) Parameter $ (z,\theta)$ for $B \rightarrow \pi \nu _{\alpha} \overline{\nu} _{\alpha},  \alpha$  are  (a) UnPolarized (b) e (c) $\mu (d) \tau$. The three contours belong to branching fraction at $[1.4-9.8]$}
\end{figure}%
%EndExpansion

%TCIMACRO{%
%\TeXButton{Linear Plot 2}{\begin{figure}[H]
%\subfigure[] {\includegraphics[scale=0.26]{Graph5.png}}
%\subfigure[] {\includegraphics[scale=0.26]{Graph6.png}}
%\subfigure[] {\includegraphics[scale=0.26]{Graph7.png}}
%\subfigure[] {\includegraphics[scale=0.26]{Graph8.png}}
%\caption{Variation of branching fraction w.r.t.NP Parameter ($\left|z(\lambda '_{i\text{jk}} \lambda '_{l\text{mn}}\right)|$,$\theta)$ for $B \rightarrow \pi \nu _{\alpha} \overline{\nu} _{\alpha},  \alpha$  is  (a) UnPolarized (b) e (c) $\mu (d) \tau$.  Experimental bound on the process is (9.8$\times10^{-5}$).}
%\end{figure}}}%
%BeginExpansion
\begin{figure}[H]
\subfigure[] {\includegraphics[scale=0.26]{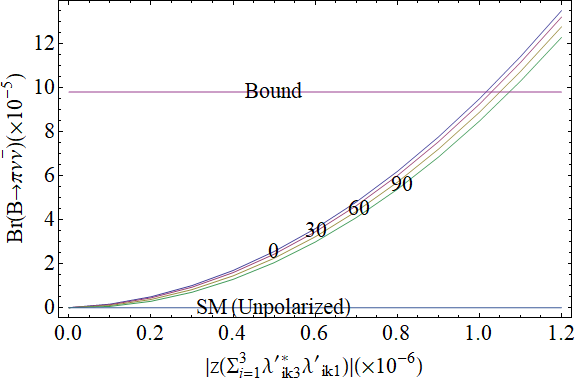}}
\subfigure[] {\includegraphics[scale=0.26]{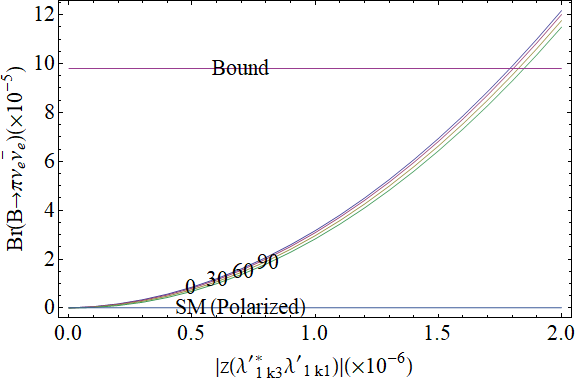}}
\subfigure[] {\includegraphics[scale=0.26]{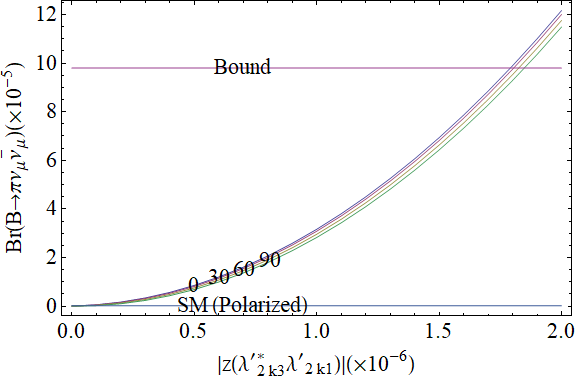}}
\subfigure[] {\includegraphics[scale=0.26]{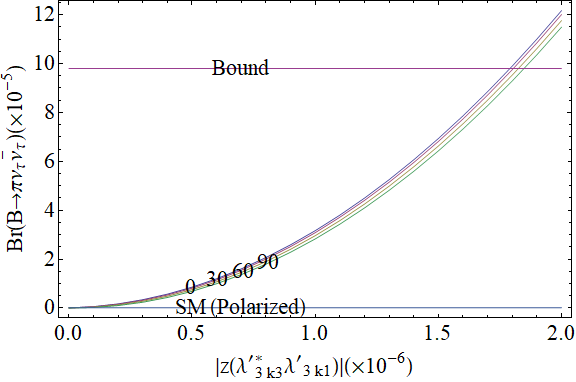}}
\caption{Variation of branching fraction w.r.t.NP Parameter ($\left|z(\lambda '_{i\text{jk}} \lambda '_{l\text{mn}}\right)|$,$\theta)$ for $B \rightarrow \pi \nu _{\alpha} \overline{\nu} _{\alpha},  \alpha$  is  (a) UnPolarized (b) e (c) $\mu (d) \tau$.  Experimental bound on the process is (9.8$\times10^{-5}$).}
\end{figure}%
%EndExpansion

%TCIMACRO{%
%\TeXButton{Bound Table 3}{\begin{table}[H]
%\begin{center}
%\subfigure[] {\includegraphics[scale=0.25]{Picture11.png}}
%\subfigure[] {\includegraphics[scale=0.25]{Picture12.png}}
%\subfigure[] {\includegraphics[scale=0.25]{Picture13.png}}
%\subfigure[] {\includegraphics[scale=0.25]{Picture14.png}}
%\end{center}
%\caption{ Bounds on NP Parameter ($\left|z(\lambda '_{i\text{jk}} \lambda '_{l\text{mn}}\right)|$,$\theta)$ for $B \rightarrow K \nu _{\alpha} \overline{\nu} _{\alpha},  \alpha$ ($Br_{SM}$)  are  (a) UnPolarized  (4.69$\times10^{-6}$) and (1.56$\times10^{-6}$) for (b) e (c) $\mu (d) \tau$. Experimental bound on the process is (1.60$\times10^{-5}$)}
%\end{table}}}%
%BeginExpansion
\begin{table}[H]
\begin{center}
\subfigure[] {\includegraphics[scale=0.25]{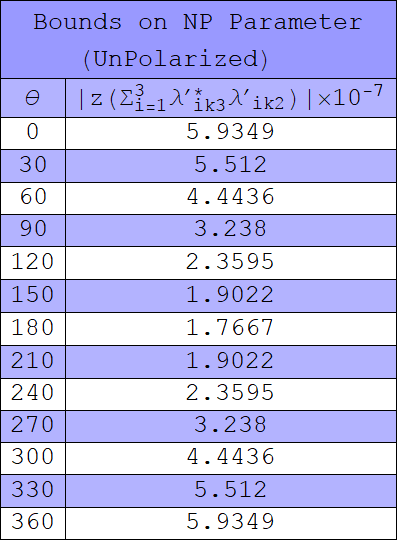}}
\subfigure[] {\includegraphics[scale=0.25]{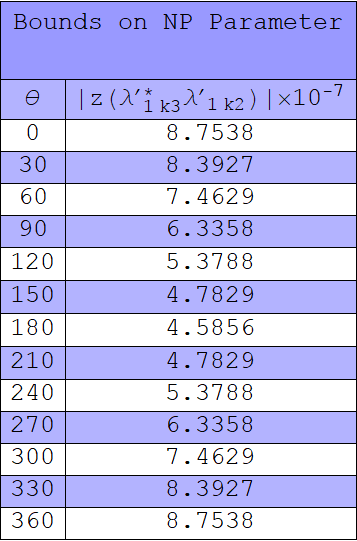}}
\subfigure[] {\includegraphics[scale=0.25]{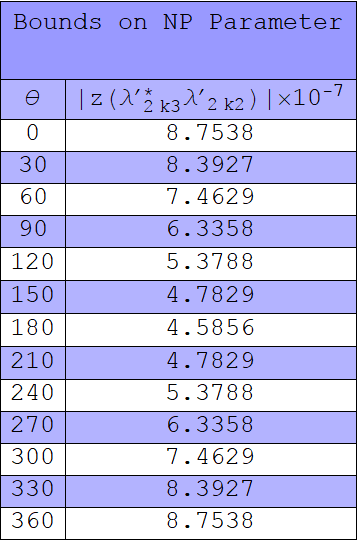}}
\subfigure[] {\includegraphics[scale=0.25]{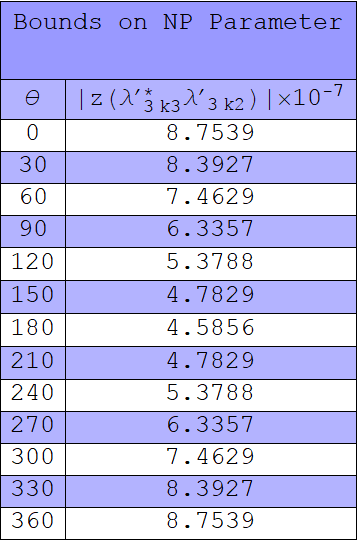}}
\end{center}
\caption{ Bounds on NP Parameter ($\left|z(\lambda '_{i\text{jk}} \lambda '_{l\text{mn}}\right)|$,$\theta)$ for $B \rightarrow K \nu _{\alpha} \overline{\nu} _{\alpha},  \alpha$ ($Br_{SM}$)  are  (a) UnPolarized  (4.69$\times10^{-6}$) and (1.56$\times10^{-6}$) for (b) e (c) $\mu (d) \tau$. Experimental bound on the process is (1.60$\times10^{-5}$)}
\end{table}%
%EndExpansion

%TCIMACRO{%
%\TeXButton{Contour Plot 4}{\begin{figure}[H]
%\begin{center}
%\subfigure[] {\includegraphics[scale=0.26]{D1.png}}
%\subfigure[] {\includegraphics[scale=0.26]{D2.png}}
%\subfigure[] {\includegraphics[scale=0.26]{D3.png}}
%\subfigure[] {\includegraphics[scale=0.26]{D4.png}}
%\end{center}
%\caption{Allowed regions of general New Physics(NP) Parameter $ (z,\theta)$ for $B \rightarrow K \nu _{\alpha} \overline{\nu} _{\alpha},  \alpha$  are  (a) UnPolarized (b) e (c) $\mu (d) \tau$. The three contours belong to branching fraction at $[0.4-1.6]$}
%\end{figure}}}%
%BeginExpansion
\begin{figure}[H]
\begin{center}
\subfigure[] {\includegraphics[scale=0.26]{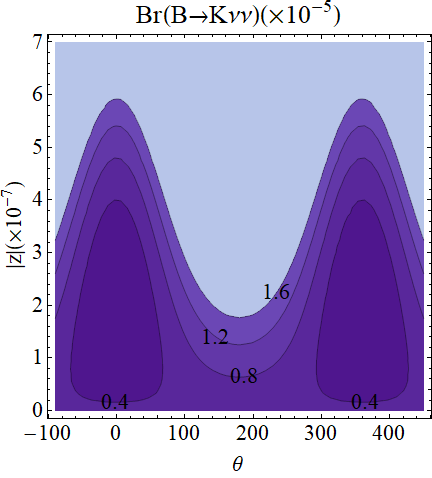}}
\subfigure[] {\includegraphics[scale=0.26]{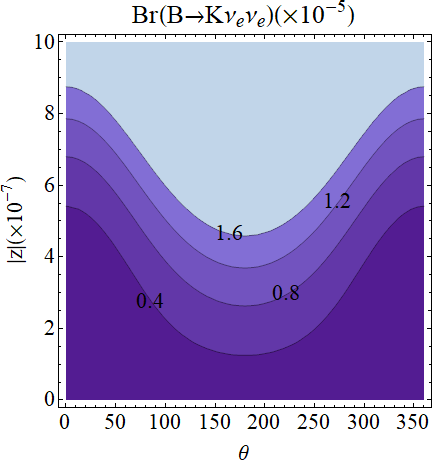}}
\subfigure[] {\includegraphics[scale=0.26]{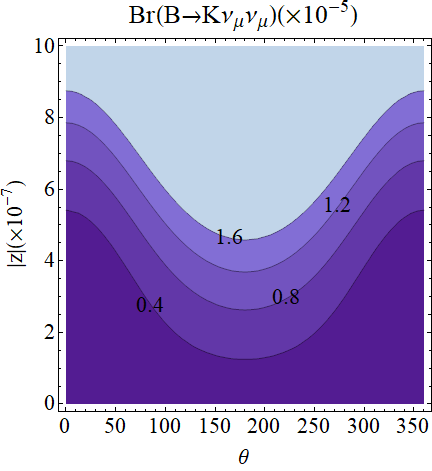}}
\subfigure[] {\includegraphics[scale=0.26]{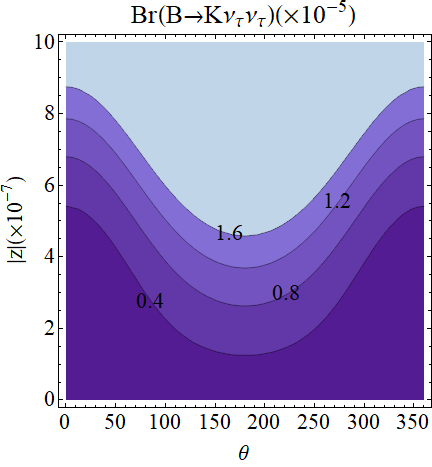}}
\end{center}
\caption{Allowed regions of general New Physics(NP) Parameter $ (z,\theta)$ for $B \rightarrow K \nu _{\alpha} \overline{\nu} _{\alpha},  \alpha$  are  (a) UnPolarized (b) e (c) $\mu (d) \tau$. The three contours belong to branching fraction at $[0.4-1.6]$}
\end{figure}%
%EndExpansion

%TCIMACRO{%
%\TeXButton{Linear Plot 3}{\begin{figure}[H]
%\begin{center}
%\subfigure[] {\includegraphics[scale=0.26]{Graph21.png}}
%\subfigure[] {\includegraphics[scale=0.26]{Graph22.png}}
%\subfigure[] {\includegraphics[scale=0.26]{Graph23.png}}
%\subfigure[] {\includegraphics[scale=0.26]{Graph24.png}}
%\end{center}
%\caption{Variations of branching fraction w.r.t.NP Parameter ($\left|z(\lambda '_{i\text{jk}} \lambda '_{l\text{mn}}\right)|$,$\theta)$ for $B \rightarrow K \nu _{\alpha} \overline{\nu} _{\alpha},  \alpha$  are  (a) UnPolarized (b) e (c) $\mu (d) \tau$.  Experimental bound on the process is (1.60$\times10^{-5}$)}
%\end{figure}}}%
%BeginExpansion
\begin{figure}[H]
\begin{center}
\subfigure[] {\includegraphics[scale=0.26]{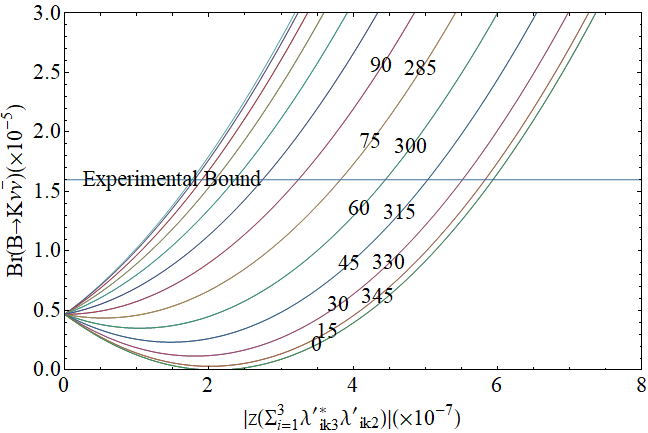}}
\subfigure[] {\includegraphics[scale=0.26]{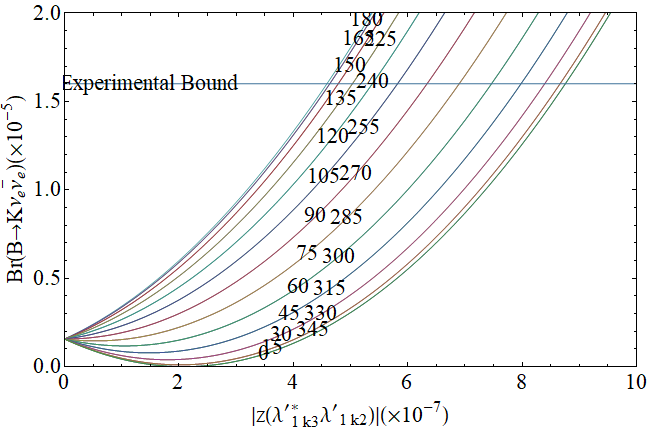}}
\subfigure[] {\includegraphics[scale=0.26]{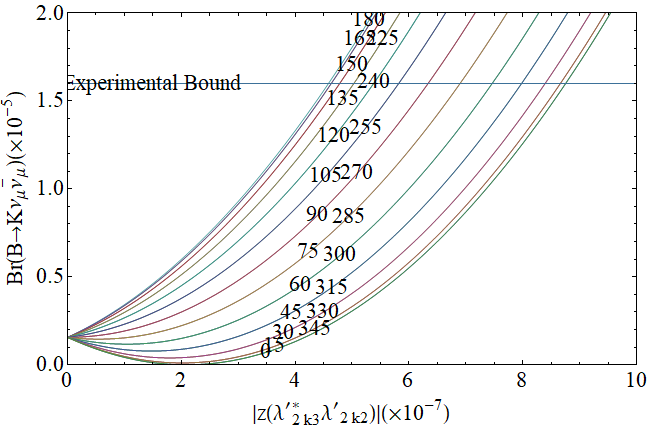}}
\subfigure[] {\includegraphics[scale=0.26]{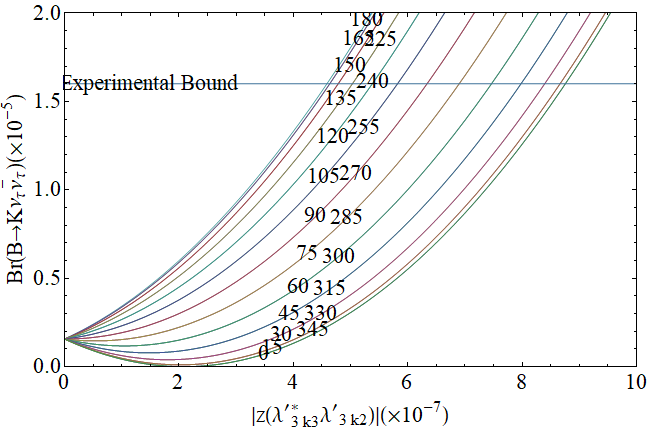}}
\end{center}
\caption{Variations of branching fraction w.r.t.NP Parameter ($\left|z(\lambda '_{i\text{jk}} \lambda '_{l\text{mn}}\right)|$,$\theta)$ for $B \rightarrow K \nu _{\alpha} \overline{\nu} _{\alpha},  \alpha$  are  (a) UnPolarized (b) e (c) $\mu (d) \tau$.  Experimental bound on the process is (1.60$\times10^{-5}$)}
\end{figure}%
%EndExpansion

%TCIMACRO{%
%\TeXButton{Bound Table 4}{\begin{table}[H]
%\subfigure[] {\includegraphics[scale=0.21]{Picture71.png}}
%\subfigure[] {\includegraphics[scale=0.21]{Picture72.png}}
%\subfigure[] {\includegraphics[scale=0.21]{Picture73.png}}
%\subfigure[] {\includegraphics[scale=0.21]{Picture74.png}}
%\caption{ Bounds on NP Parameter (derived from $K \rightarrow \pi \nu _{\alpha} \overline{\nu} _{\alpha},  \alpha$)  ($\left|\lambda '_{i\text{jk}} \lambda '_{l\text{mn}}\right|$,$\theta)$for $K_{s} \rightarrow \nu_{\alpha} \overline{\nu} _{\alpha},  \alpha$ ($Br_{SM}$)  are  (a) UnPolarized  (7.27$\times10^{-27})$(b) e  (7.58$\times10^{-33})$ (c) $\mu$  (5.66$\times10^{-29})$ (d) $\tau$ (7.27$\times10^{-27})$.}
%\end{table}}}%
%BeginExpansion
\begin{table}[H]
\subfigure[] {\includegraphics[scale=0.21]{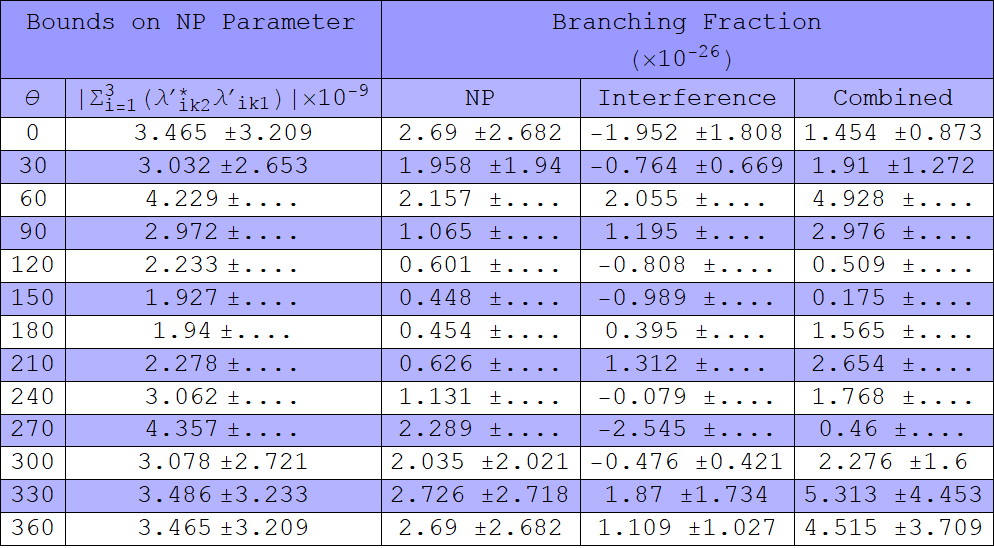}}
\subfigure[] {\includegraphics[scale=0.21]{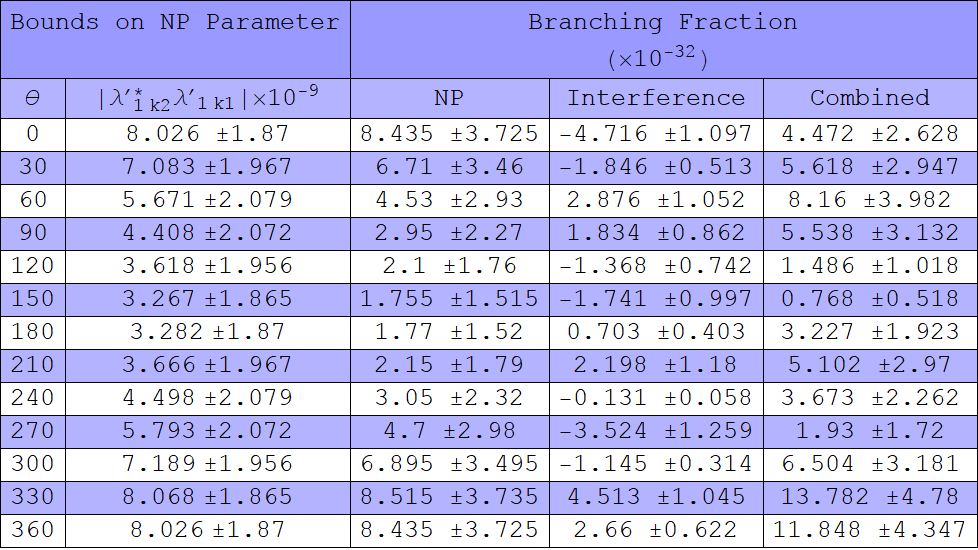}}
\subfigure[] {\includegraphics[scale=0.21]{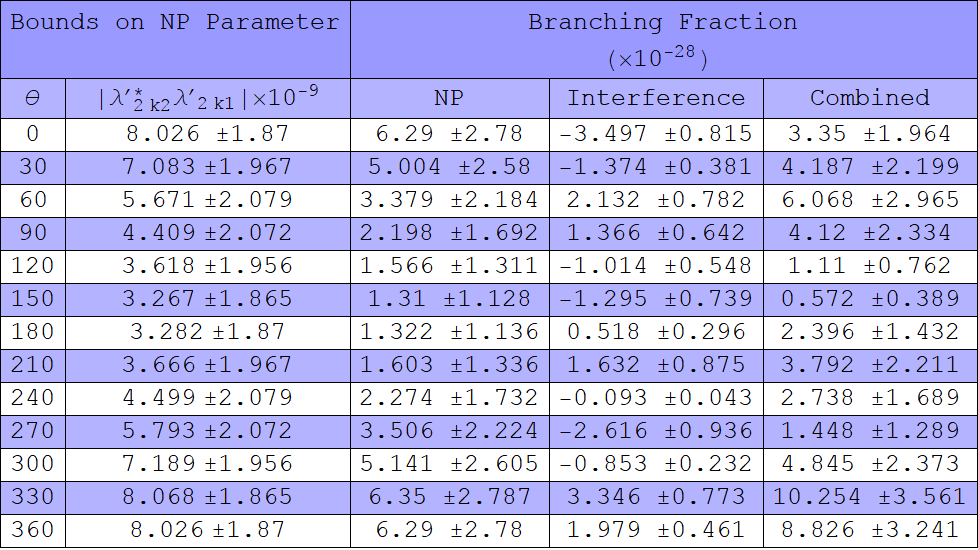}}
\subfigure[] {\includegraphics[scale=0.21]{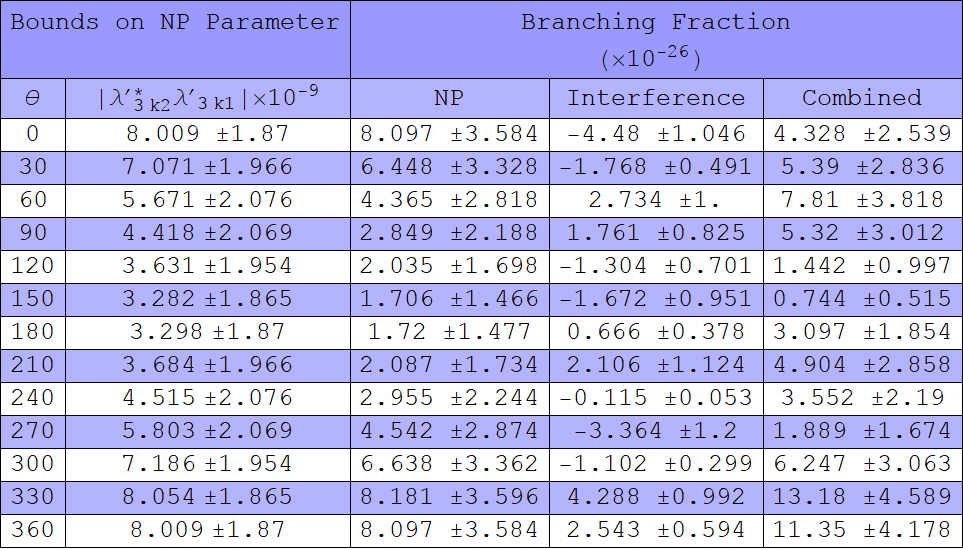}}
\caption{ Bounds on NP Parameter (derived from $K \rightarrow \pi \nu _{\alpha} \overline{\nu} _{\alpha},  \alpha$)  ($\left|\lambda '_{i\text{jk}} \lambda '_{l\text{mn}}\right|$,$\theta)$for $K_{s} \rightarrow \nu_{\alpha} \overline{\nu} _{\alpha},  \alpha$ ($Br_{SM}$)  are  (a) UnPolarized  (7.27$\times10^{-27})$(b) e  (7.58$\times10^{-33})$ (c) $\mu$  (5.66$\times10^{-29})$ (d) $\tau$ (7.27$\times10^{-27})$.}
\end{table}%
%EndExpansion

%TCIMACRO{%
%\TeXButton{Linear Plot 4}{\begin{figure}[H]
%\begin{center}
%\subfigure[] {\includegraphics[scale=0.26]{NPKS1.png}}
%\subfigure[] {\includegraphics[scale=0.26]{NPKS2.png}}
%\subfigure[] {\includegraphics[scale=0.26]{NPKS3.png}}
%\subfigure[] {\includegraphics[scale=0.26]{NPKS4.png}}
%\end{center}
%\caption{Variations of NP contribution w.r.t.NP Parameter ($\left|\lambda '_{i\text{jk}} \lambda '_{l\text{mn}}\right|$,$\theta)$ for $K _S \rightarrow  \nu _{\alpha} \overline{\nu} _{\alpha},  \alpha$  are  (a) UnPolarized (b) e (c) $\mu (d) \tau$.}
%\end{figure}}}%
%BeginExpansion
\begin{figure}[H]
\begin{center}
\subfigure[] {\includegraphics[scale=0.26]{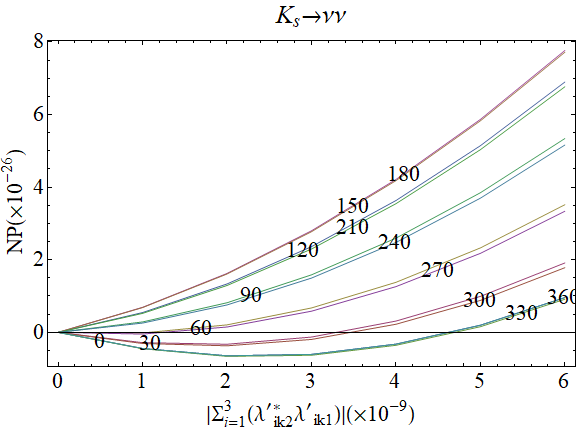}}
\subfigure[] {\includegraphics[scale=0.26]{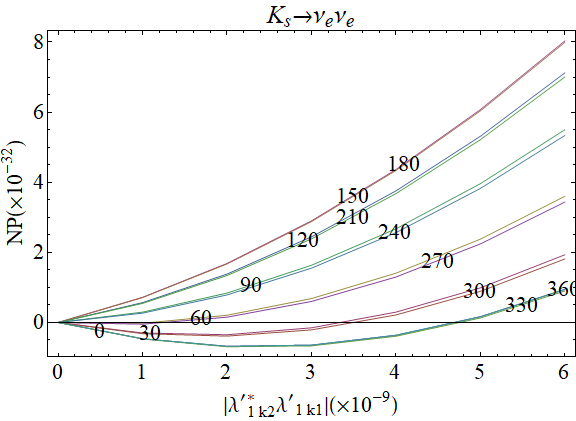}}
\subfigure[] {\includegraphics[scale=0.26]{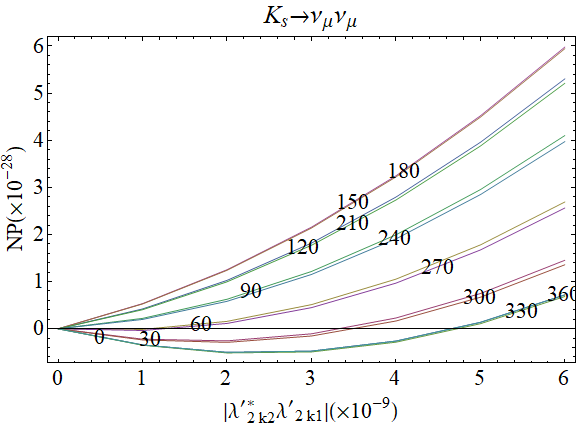}}
\subfigure[] {\includegraphics[scale=0.26]{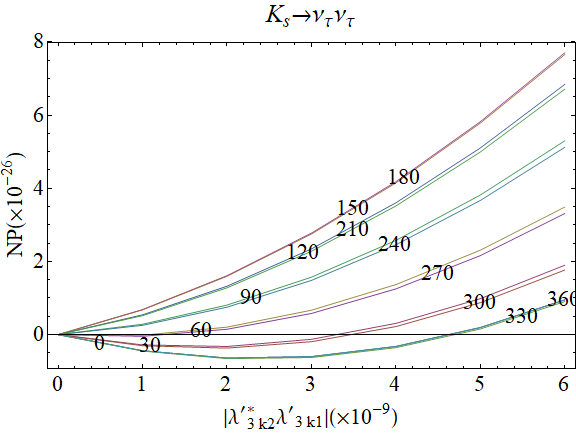}}
\end{center}
\caption{Variations of NP contribution w.r.t.NP Parameter ($\left|\lambda '_{i\text{jk}} \lambda '_{l\text{mn}}\right|$,$\theta)$ for $K _S \rightarrow  \nu _{\alpha} \overline{\nu} _{\alpha},  \alpha$  are  (a) UnPolarized (b) e (c) $\mu (d) \tau$.}
\end{figure}%
%EndExpansion

%TCIMACRO{%
%\TeXButton{Bound Table 5}{\begin{table}[H]
%\subfigure[] {\includegraphics[scale=0.19]{Picture75.png}}
%\subfigure[] {\includegraphics[scale=0.21]{Picture76.png}}
%\subfigure[] {\includegraphics[scale=0.21]{Picture77.png}}
%\subfigure[] {\includegraphics[scale=0.21]{Picture78.png}}
%\caption{ Bounds on NP Parameter (derived from $K \rightarrow \pi \nu _{\alpha} \overline{\nu} _{\alpha},  \alpha$) ($\left|\lambda '_{i\text{jk}} \lambda '_{l\text{mn}}\right|$,$\theta)$ for $K_{L} \rightarrow \nu_{\alpha} \overline{\nu} _{\alpha},  \alpha$ ($Br_{SM}$) are  (a) UnPolarized (4.29$\times10^{-27})$ (b) e (4.48$\times10^{-33})$  (c) $\mu$  (3.34$\times10^{-29})$(d) $\tau$ (4.26$\times10^{-27})$.}
%\end{table}}}%
%BeginExpansion
\begin{table}[H]
\subfigure[] {\includegraphics[scale=0.19]{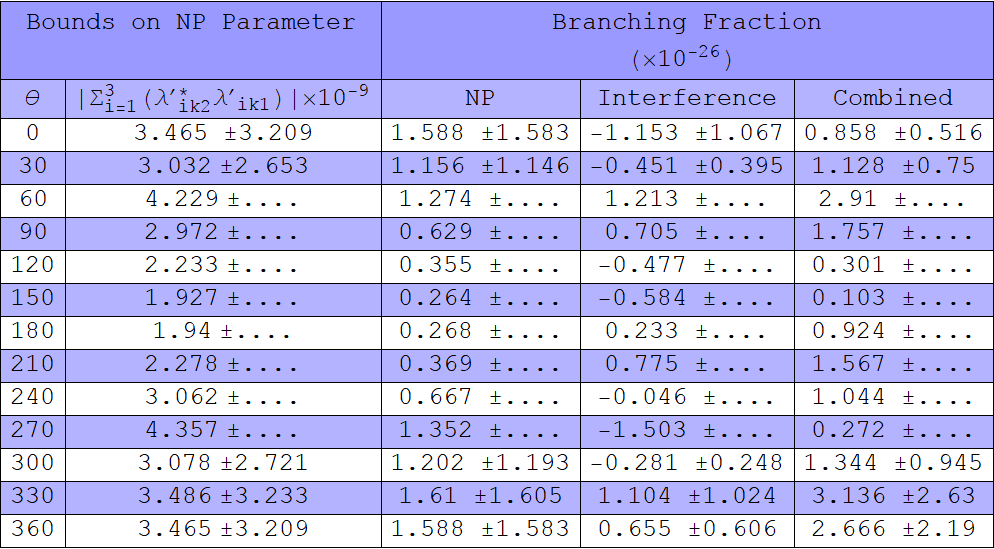}}
\subfigure[] {\includegraphics[scale=0.21]{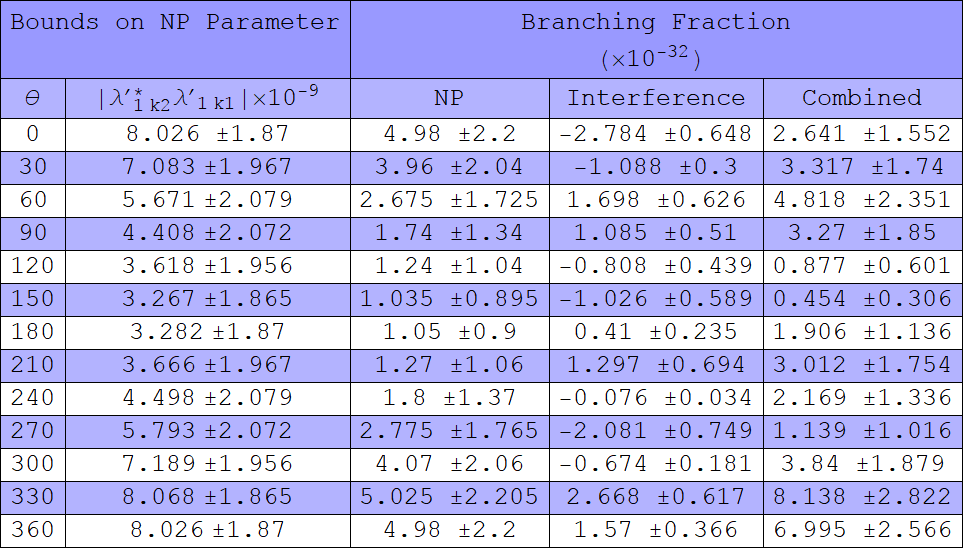}}
\subfigure[] {\includegraphics[scale=0.21]{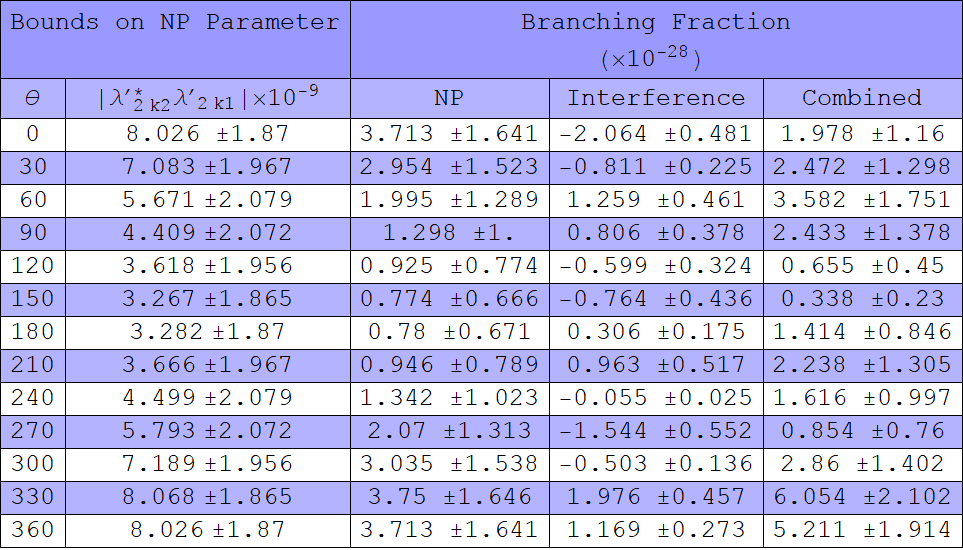}}
\subfigure[] {\includegraphics[scale=0.21]{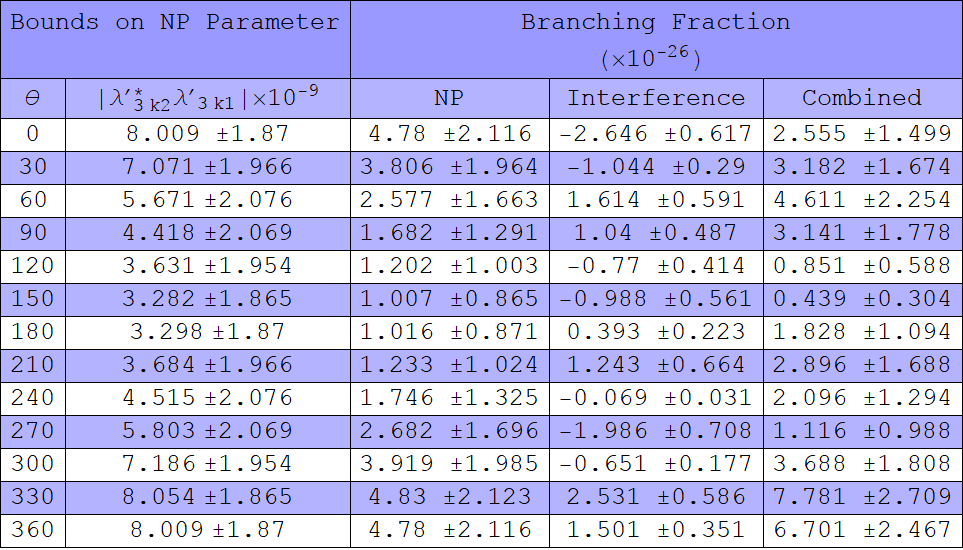}}
\caption{ Bounds on NP Parameter (derived from $K \rightarrow \pi \nu _{\alpha} \overline{\nu} _{\alpha},  \alpha$) ($\left|\lambda '_{i\text{jk}} \lambda '_{l\text{mn}}\right|$,$\theta)$ for $K_{L} \rightarrow \nu_{\alpha} \overline{\nu} _{\alpha},  \alpha$ ($Br_{SM}$) are  (a) UnPolarized (4.29$\times10^{-27})$ (b) e (4.48$\times10^{-33})$  (c) $\mu$  (3.34$\times10^{-29})$(d) $\tau$ (4.26$\times10^{-27})$.}
\end{table}%
%EndExpansion

%TCIMACRO{%
%\TeXButton{Linear Plot5}{\begin{figure}[H]
%\begin{center}
%\subfigure[] {\includegraphics[scale=0.26]{NPKL1.png}}
%\subfigure[] {\includegraphics[scale=0.26]{NPKL2.png}}
%\subfigure[] {\includegraphics[scale=0.26]{NPKL3.png}}
%\subfigure[] {\includegraphics[scale=0.26]{NPKL4.png}}
%\end{center}
%\caption{Variations of NP contribution w.r.t.NP Parameter ($\left|\lambda '_{i\text{jk}} \lambda '_{l\text{mn}}\right|$,$\theta)$ for $K _L \rightarrow  \nu _{\alpha} \overline{\nu} _{\alpha},  \alpha$  are  (a) UnPolarized (b) e (c) $\mu (d) \tau$.}
%\end{figure}}}%
%BeginExpansion
\begin{figure}[H]
\begin{center}
\subfigure[] {\includegraphics[scale=0.26]{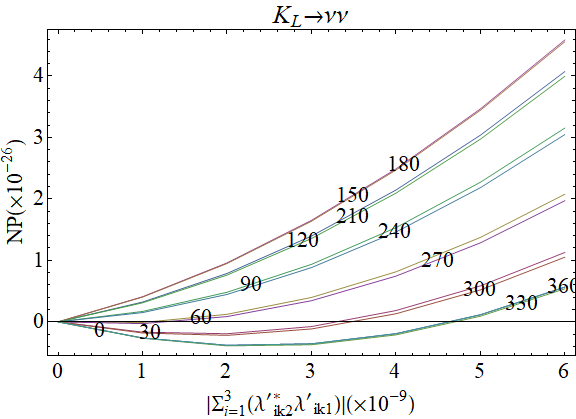}}
\subfigure[] {\includegraphics[scale=0.26]{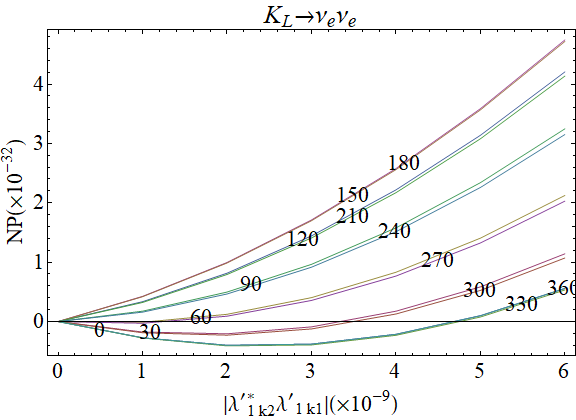}}
\subfigure[] {\includegraphics[scale=0.26]{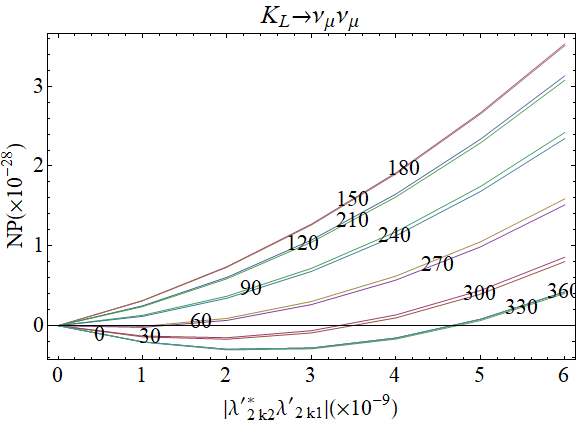}}
\subfigure[] {\includegraphics[scale=0.26]{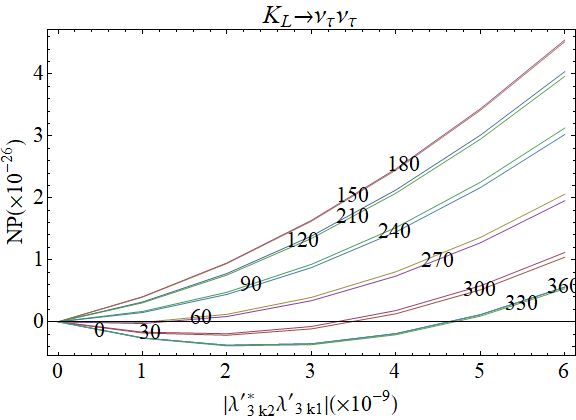}}
\end{center}
\caption{Variations of NP contribution w.r.t.NP Parameter ($\left|\lambda '_{i\text{jk}} \lambda '_{l\text{mn}}\right|$,$\theta)$ for $K _L \rightarrow  \nu _{\alpha} \overline{\nu} _{\alpha},  \alpha$  are  (a) UnPolarized (b) e (c) $\mu (d) \tau$.}
\end{figure}%
%EndExpansion

%TCIMACRO{%
%\TeXButton{Bound Table 6}{\begin{table}[H]
%\subfigure[] {\includegraphics[scale=0.22]{Picture781.png}}
%\subfigure[] {\includegraphics[scale=0.22]{Picture79.png}}
%\subfigure[] {\includegraphics[scale=0.22]{Picture80.png}}
%\subfigure[] {\includegraphics[scale=0.22]{Picture81.png}}
%\caption{ Bounds on NP Parameter (derived from $B \rightarrow \pi \nu _{\alpha} \overline{\nu} _{\alpha},  \alpha$)  ($\left|\lambda '_{i\text{jk}} \lambda '_{l\text{mn}}\right|$,$\theta)$ for $B_{d} \rightarrow \nu_{\alpha} \overline{\nu} _{\alpha},  \alpha$ ($Br_{SM}$)  are  (a) UnPolarized (6.35$\times10^{-25})$ (b) e (6.53$\times10^{-31})$ (c) $\mu$ (4.87$\times10^{-27}$) (d)$\tau$ (6.3$\times10^{-25}$).}
%\end{table}}}%
%BeginExpansion
\begin{table}[H]
\subfigure[] {\includegraphics[scale=0.22]{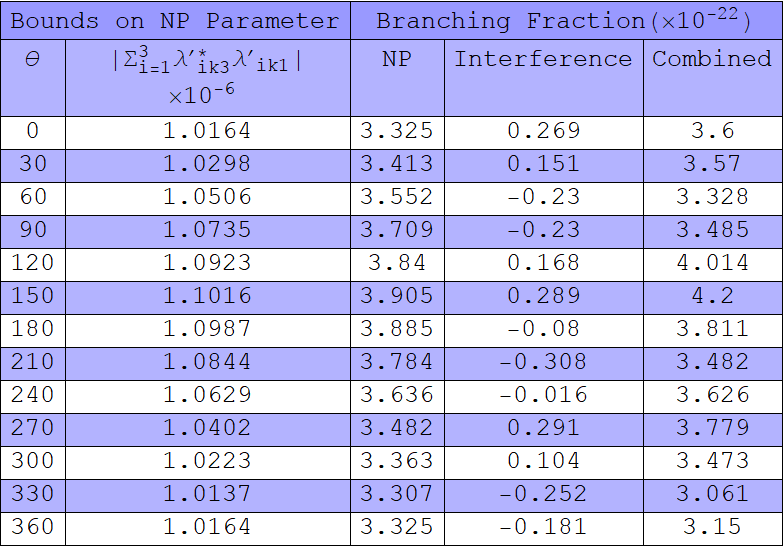}}
\subfigure[] {\includegraphics[scale=0.22]{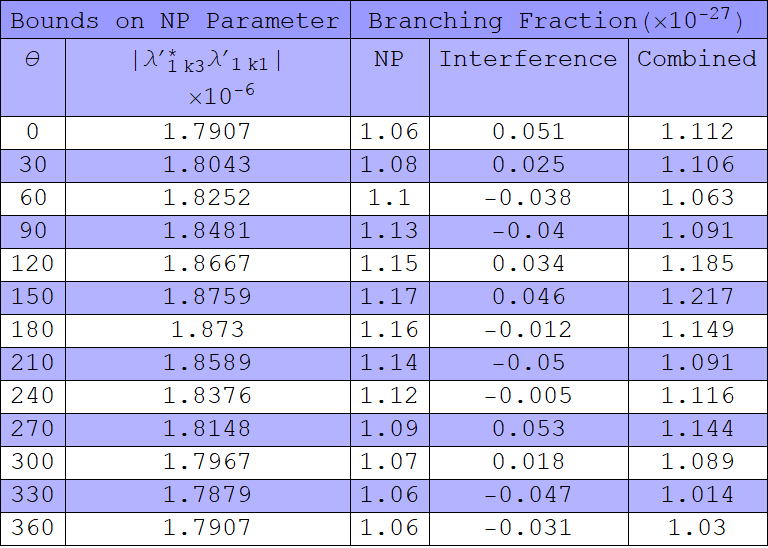}}
\subfigure[] {\includegraphics[scale=0.22]{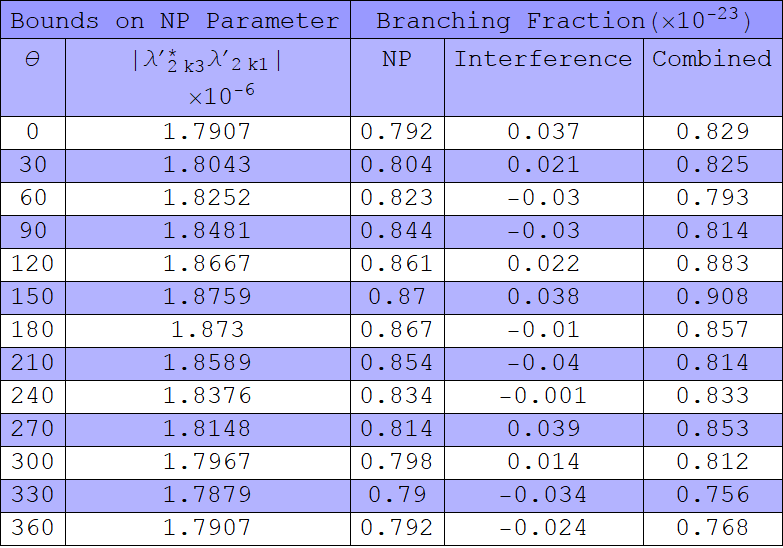}}
\subfigure[] {\includegraphics[scale=0.22]{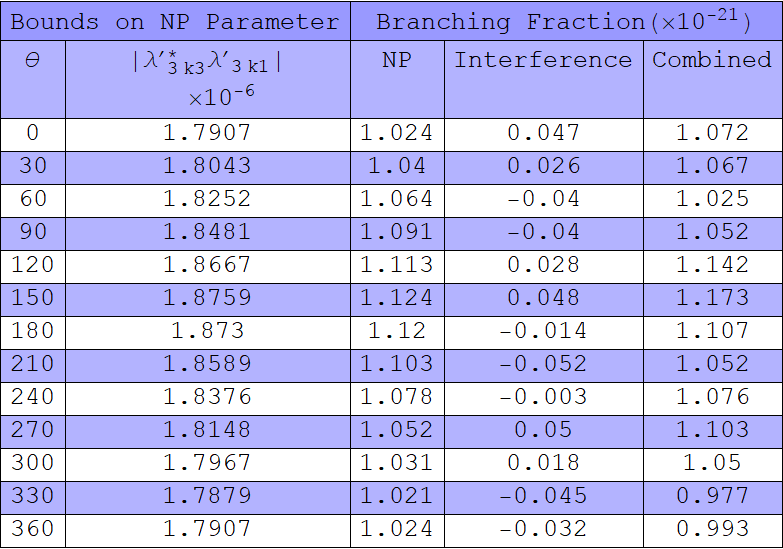}}
\caption{ Bounds on NP Parameter (derived from $B \rightarrow \pi \nu _{\alpha} \overline{\nu} _{\alpha},  \alpha$)  ($\left|\lambda '_{i\text{jk}} \lambda '_{l\text{mn}}\right|$,$\theta)$ for $B_{d} \rightarrow \nu_{\alpha} \overline{\nu} _{\alpha},  \alpha$ ($Br_{SM}$)  are  (a) UnPolarized (6.35$\times10^{-25})$ (b) e (6.53$\times10^{-31})$ (c) $\mu$ (4.87$\times10^{-27}$) (d)$\tau$ (6.3$\times10^{-25}$).}
\end{table}%
%EndExpansion

%TCIMACRO{%
%\TeXButton{Linear Plot 6}{\begin{figure}[H]
%\begin{center}
%\subfigure[] {\includegraphics[scale=0.26]{NPBD1.png}}
%\subfigure[] {\includegraphics[scale=0.26]{NPBD2.png}}
%\subfigure[] {\includegraphics[scale=0.26]{NPBD3.png}}
%\subfigure[] {\includegraphics[scale=0.26]{NPBD4.png}}
%\end{center}
%\caption{Variations of NP contribution w.r.t.NP Parameter ($\left|\lambda '_{i\text{jk}} \lambda '_{l\text{mn}}\right|$,$\theta)$ for $B _d \rightarrow  \nu _{\alpha} \overline{\nu} _{\alpha},  \alpha$  are  (a) UnPolarized (b) e (c) $\mu (d) \tau$.}
%\end{figure}}}%
%BeginExpansion
\begin{figure}[H]
\begin{center}
\subfigure[] {\includegraphics[scale=0.26]{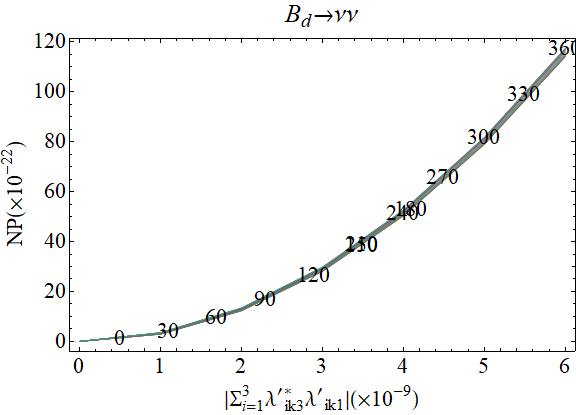}}
\subfigure[] {\includegraphics[scale=0.26]{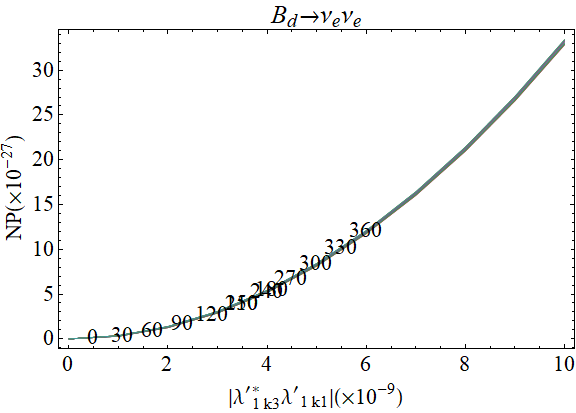}}
\subfigure[] {\includegraphics[scale=0.26]{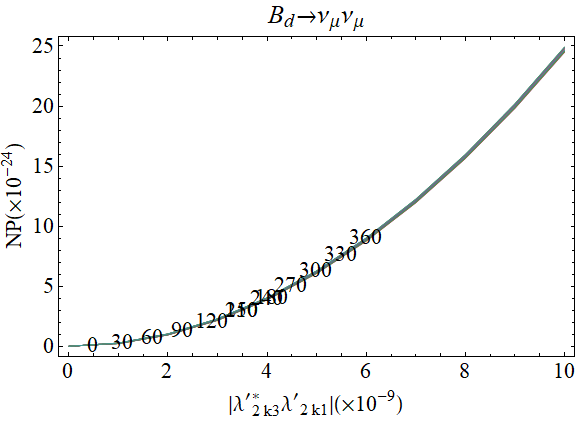}}
\subfigure[] {\includegraphics[scale=0.26]{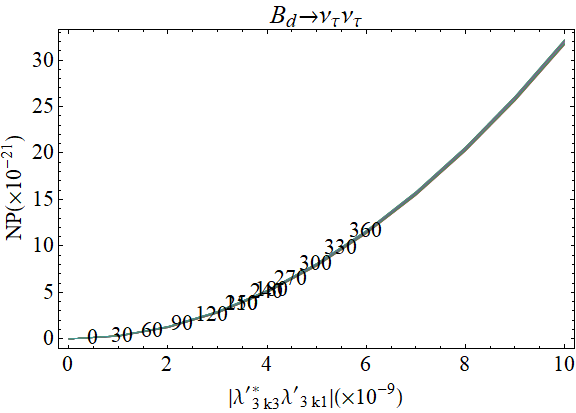}}
\end{center}
\caption{Variations of NP contribution w.r.t.NP Parameter ($\left|\lambda '_{i\text{jk}} \lambda '_{l\text{mn}}\right|$,$\theta)$ for $B _d \rightarrow  \nu _{\alpha} \overline{\nu} _{\alpha},  \alpha$  are  (a) UnPolarized (b) e (c) $\mu (d) \tau$.}
\end{figure}%
%EndExpansion

%TCIMACRO{%
%\TeXButton{Bound Table 7}{\begin{table}[H]
%\subfigure[] {\includegraphics[scale=0.22]{Picture64.png}}
%\subfigure[] {\includegraphics[scale=0.22]{Picture65.png}}
%\subfigure[] {\includegraphics[scale=0.22]{Picture66.png}}
%\subfigure[] {\includegraphics[scale=0.22]{Picture67.png}}
%\caption{ Bounds on NP Parameter (derived from $B \rightarrow K \nu _{\alpha} \overline{\nu} _{\alpha},  \alpha$)  ($\left|\lambda '_{i\text{jk}} \lambda '_{l\text{mn}}\right|$,$\theta)$ for $B_{s} \rightarrow \nu_{\alpha} \overline{\nu} _{\alpha},  \alpha$ ($Br_{SM}$)  are  (a) UnPolarized (1.41$\times10^{-23})$ (b) e (1.45$\times10^{-29})$ (c) $\mu$(1.08$\times10^{-25}$) (d)$\tau$(1.4$\times10^{-23}$).}
%\end{table}}}%
%BeginExpansion
\begin{table}[H]
\subfigure[] {\includegraphics[scale=0.22]{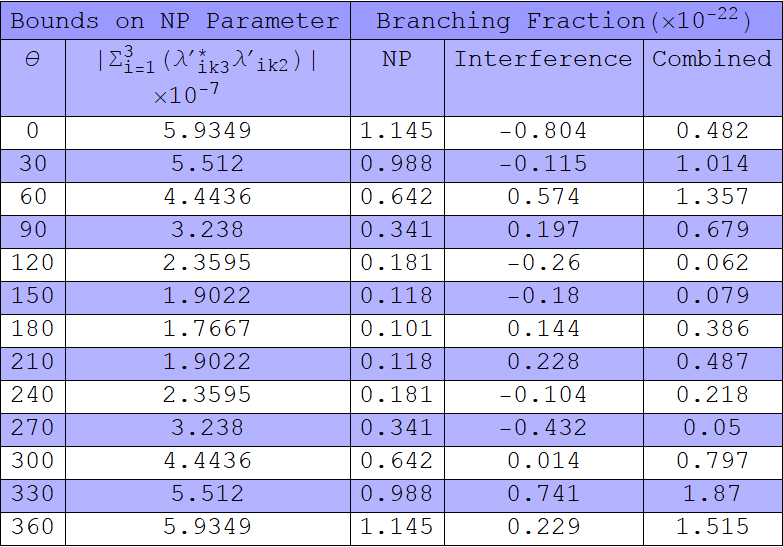}}
\subfigure[] {\includegraphics[scale=0.22]{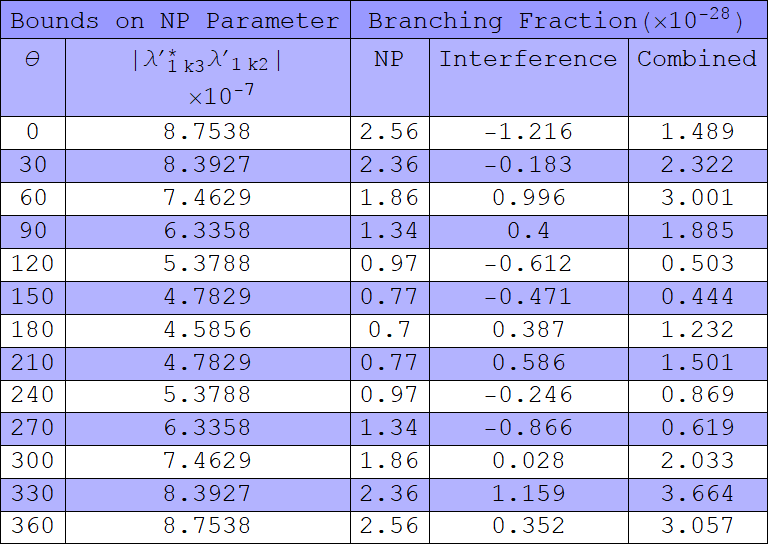}}
\subfigure[] {\includegraphics[scale=0.22]{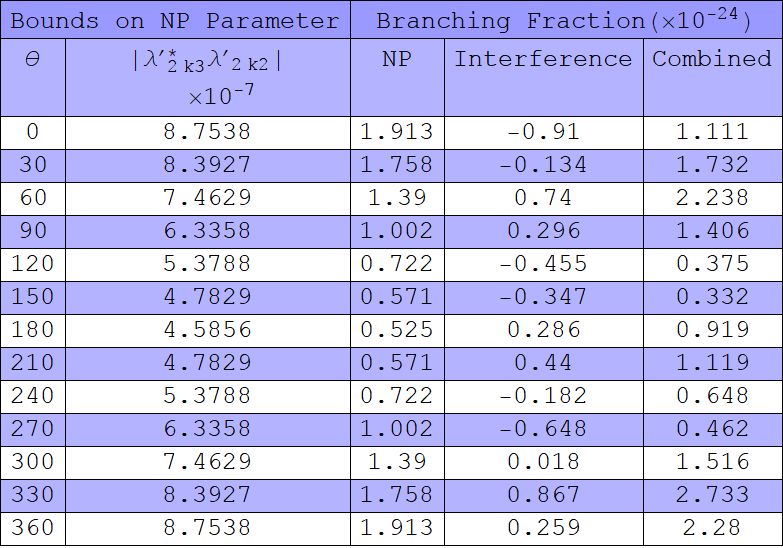}}
\subfigure[] {\includegraphics[scale=0.22]{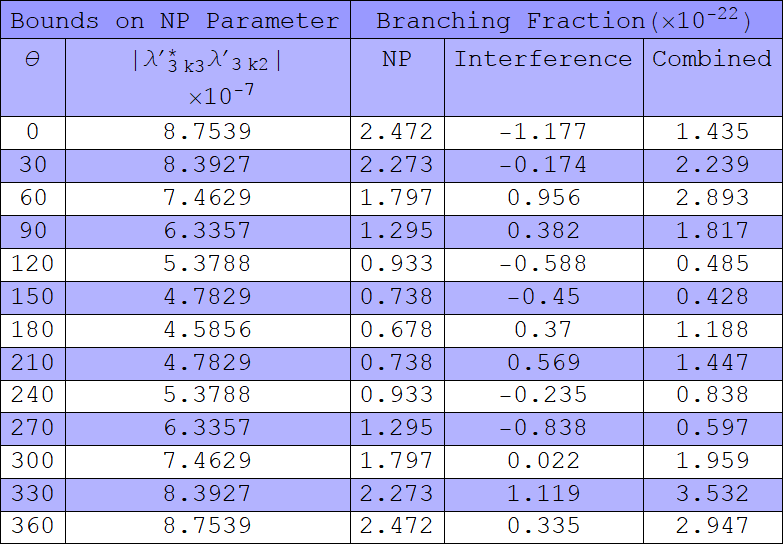}}
\caption{ Bounds on NP Parameter (derived from $B \rightarrow K \nu _{\alpha} \overline{\nu} _{\alpha},  \alpha$)  ($\left|\lambda '_{i\text{jk}} \lambda '_{l\text{mn}}\right|$,$\theta)$ for $B_{s} \rightarrow \nu_{\alpha} \overline{\nu} _{\alpha},  \alpha$ ($Br_{SM}$)  are  (a) UnPolarized (1.41$\times10^{-23})$ (b) e (1.45$\times10^{-29})$ (c) $\mu$(1.08$\times10^{-25}$) (d)$\tau$(1.4$\times10^{-23}$).}
\end{table}%
%EndExpansion

%TCIMACRO{%
%\TeXButton{Linear Plot 7}{\begin{figure}[H]
%\begin{center}
%\subfigure[] {\includegraphics[scale=0.26]{NPBS1.png}}
%\subfigure[] {\includegraphics[scale=0.26]{NPBS2.png}}
%\subfigure[] {\includegraphics[scale=0.26]{NPBS3.png}}
%\subfigure[] {\includegraphics[scale=0.26]{NPBS4.png}}
%\end{center}
%\caption{Variations of NP contribution w.r.t.NP Parameter ($\left|\lambda '_{i\text{jk}} \lambda '_{l\text{mn}}\right|$,$\theta)$ for $B _s \rightarrow  \nu _{\alpha} \overline{\nu} _{\alpha},  \alpha$  are  (a) UnPolarized (b) e (c) $\mu (d) \tau$.}
%\end{figure}}}%
%BeginExpansion
\begin{figure}[H]
\begin{center}
\subfigure[] {\includegraphics[scale=0.26]{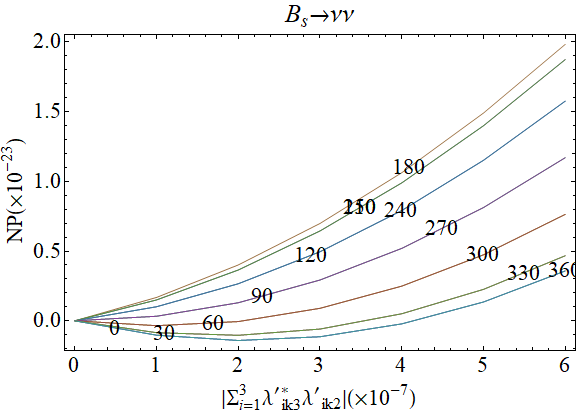}}
\subfigure[] {\includegraphics[scale=0.26]{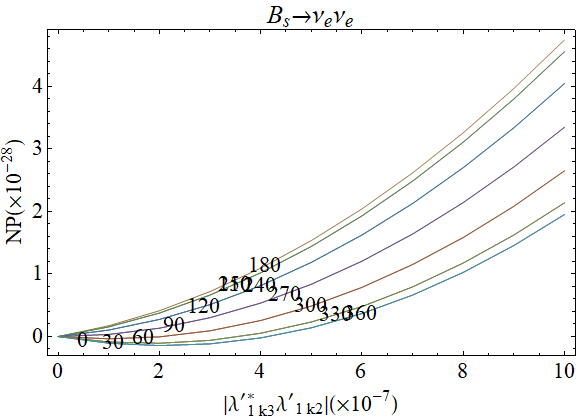}}
\subfigure[] {\includegraphics[scale=0.26]{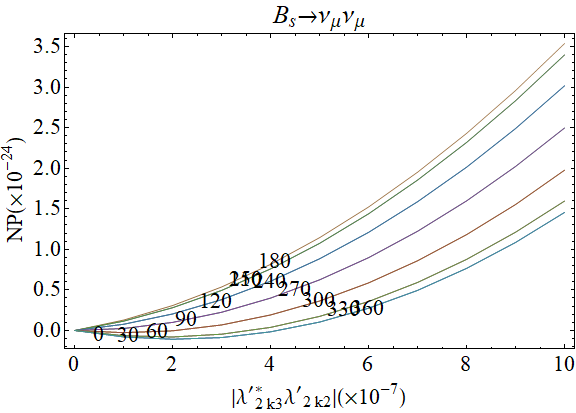}}
\subfigure[] {\includegraphics[scale=0.26]{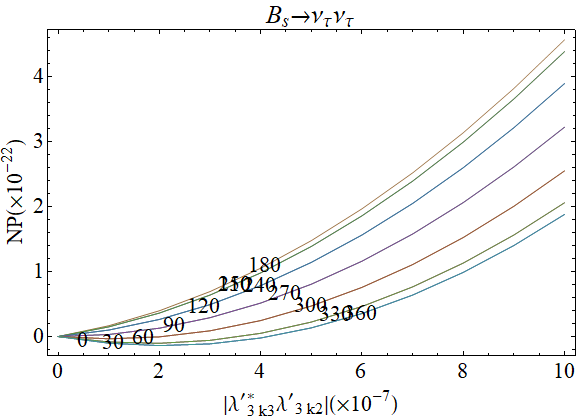}}
\end{center}
\caption{Variations of NP contribution w.r.t.NP Parameter ($\left|\lambda '_{i\text{jk}} \lambda '_{l\text{mn}}\right|$,$\theta)$ for $B _s \rightarrow  \nu _{\alpha} \overline{\nu} _{\alpha},  \alpha$  are  (a) UnPolarized (b) e (c) $\mu (d) \tau$.}
\end{figure}%
%EndExpansion

Summarizing, we have carried out the study of semileptonic and pure leptonic
decays of pseudoscalar mesons within $\NEG{R}_{p}$ MSSM. It enhances the SM
contribution for all the involved processes as discussed in the context of
decays of $K^{0},~K_{S,L}$ and$~B_{s,d}$. This makes $\NEG{R}_{p}$ MSSM a
viable model for checking the contribution of NP in rare decays.

\end{document}